\documentclass[a4paper,11pt]{article}

\pdfoutput=1

\usepackage[utf8]{inputenc}
\usepackage{dsfont}
\usepackage{epsfig}
\usepackage{slashed}
\usepackage{bbold}
\usepackage{psfrag}
\usepackage{color}
\PassOptionsToPackage{caption=false}{subfig}
\usepackage{subfig}
\usepackage{multirow}
\usepackage{booktabs}
\usepackage{pstricks}
\usepackage{feynmp}
\usepackage{jheppub} 
\usepackage{cases}

\usepackage[normalem]{ulem}

\newcommand{\cF}{\cos \frac{\phi}{F}}

\newcommand{\sFo}{\sin \frac{\phi_0}{F}}
\newcommand{\cf}{\cos \frac{\phi}{f}}

\newcommand{\HD}{\mathcal{H}}
\newcommand{\Mp}{M_{\rm Pl}}
\newcommand{\dphi}{\delta \phi}

\def\beqn{\begin{eqnarray}} 
\def\eeqn{\end{eqnarray}} 
\def\be{\begin{equation}}
\def\ee{\end{equation}}
\def\nn{\nonumber}

\begin{document}
 \title{\begin{center} Dynamics of Relaxed Inflation \end{center}}

 \author[a]{Walter Tangarife,}
 \author[a, b]{Kohsaku Tobioka,}
 \author[a]{Lorenzo Ubaldi,}
 \author[a]{Tomer Volansky}

\affiliation[a]{Raymond and Beverly Sackler School of Physics and Astronomy, \\
 Tel-Aviv University, Tel-Aviv 69978, Israel}
 
\affiliation[b]{Department of Particle Physics and Astrophysics, \\
Weizmann Institute of Science, Rehovot 76100, Israel}

\emailAdd{waltert@post.tau.ac.il}
\emailAdd{kohsakut@post.tau.ac.il}
\emailAdd{ubaldi.physics@gmail.com}
\emailAdd{tomerv@post.tau.ac.il}

\abstract{
The cosmological relaxation of the electroweak scale has been proposed as a mechanism to address the hierarchy problem of the Standard Model. A field, the relaxion, rolls down its potential and, in doing so, scans the squared mass parameter of the Higgs, relaxing it to a parametrically small value. In this work, we promote the relaxion to an inflaton. We couple it to Abelian gauge bosons, thereby introducing the necessary dissipation mechanism which slows down the field in the last stages. We describe a novel reheating mechanism, which relies on the gauge-boson production leading to strong electromagnetic fields, and proceeds via the vacuum production of electron-positron pairs through the Schwinger effect. We refer to this mechanism as Schwinger reheating. We discuss the cosmological dynamics of the model and the phenomenological constraints from CMB and other experiments. We find that a cutoff close to the Planck scale may be achieved. In its minimal form, the model does not generate sufficient curvature perturbations and additional ingredients, such as a curvaton field, are needed.
}
\maketitle

\section{Introduction}
During the past few decades, numerous ideas to solve the hierarchy problem have been put forth.  The majority of these ideas require the introduction of new symmetries to protect the Higgs mass. Such symmetries lead to the prediction of new degrees of freedom at the electroweak (EW) scale, but none of them has
yet been found in experiments. In Ref.~\cite{Graham:2015cka}, a new mechanism was proposed as an alternative solution to the hierarchy problem. 
The proposal relies on an axion-like field, dubbed relaxion, coupled to the Higgs field in an effective Lagrangian. The relaxion, during its cosmological
evolution, scans the Higgs mass, and finally settles on a local minimum where the Higgs has the observed mass, which is 
parametrically smaller than the cutoff of the effective Lagrangian. The larger the cutoff, the more successful the mechanism
in addressing the hierarchy problem. Several works~\cite{Espinosa:2015eda, Hardy:2015laa, Jaeckel:2015txa, Gupta:2015uea, Batell:2015fma, Matsedonskyi:2015xta, Marzola:2015dia, Choi:2015fiu, Kaplan:2015fuy, DiChiara:2015euo, Ibanez:2015fcv, Hebecker:2015zss, Fonseca:2016eoo, Fowlie:2016jlx, Evans:2016htp, Huang:2016dhp, Kobayashi:2016bue, Hook:2016mqo, Higaki:2016cqb, Choi:2016luu, Flacke:2016szy, McAllister:2016vzi, Choi:2016kke, Lalak:2016mbv, You:2017kah, Evans:2017bjs,Batell:2017kho, Beauchesne:2017ukw} have elaborated on various aspects of this framework. 

 In the original proposal~\cite{Graham:2015cka}, the entire
scanning takes place in the background of inflation, which provides constant Hubble friction necessary for the relaxion
to maintain slow roll and eventually stop at the local minimum.  This setup is rather constraining, rendering the cutoff scale significantly below the Planck scale.  
A natural question to ask is: can some of the restrictions be ameliorated by promoting the relaxion to play the role of the inflaton itself?  
In this article, we answer this question positively, extending the discussion presented in \cite{Tangarife:2017vnd}.    
One pleasant consequence of this promotion is that we can indeed achieve a higher cutoff compared to the previous proposal.
The price to pay is that the dynamics of the relaxion require a friction mechanism that can remain efficient after the reheating process ends, preventing the overshooting of the EW minimum.  
Here we consider friction from the tachyonic production of gauge bosons, due to the time-dependent
 background of the rolling relaxion. Not only this slows down the field efficiently, 
 but also provides an interesting and novel mechanism for reheating.  
A possible issue related to such a friction mechanism is that one risks overproducing the cosmological perturbations. 
We show there is a region of parameter space where this is avoided.

The relaxion dynamics proceeds in three regimes. The first consists of a long period of inflation, with standard slow roll due to the flatness of the potential. The second corresponds to the last $\cal O$(20) e-folds of inflation, where slow roll is due to dissipation via
gauge-boson production. At this stage, reheating takes place.  The third is after reheating, where the relaxion keeps rolling, the friction from gauge-boson production 
is still present and sufficient to allow for the field to settle on the correct local minimum. Thus, the final stage of relaxation of the EW scale 
occurs {\em after} the end of inflation. This is described with detail in Sections \ref{sec:model} and \ref{sec:inflation}. The last two regimes share similarities with models of axion inflation that have been studied previously in Refs.~\cite{Anber:2009ua, Barnaby:2011vw, Linde:2012bt, Pajer:2013fsa}. The relevance of this type of dissipation, in the context of relaxion models, has been also discussed in Refs.~\cite{Hook:2016mqo, Choi:2016kke}.  

An important aspect of our work is the actual reheating mechanism, which, to the best of our knowledge, has not been explored before.\footnote{As we were finishing this work, another thermalization mechanism, involving scattering between the inflaton and the gauge bosons, 
was presented in the context of axion-inflation models \cite{Ferreira:2017lnd}.}    
The photons that are produced in the last stages of inflation have a large occupation number and very low momentum. This is a coherent collection that is best described classically as an electromagnetic field. 
The electric field within the horizon is constant, to a good approximation, and its strength can grow to very large values. Eventually,
this allows for vacuum creation of electron-positron pairs through the Schwinger mechanism~\cite{Heisenberg:1935qt,Schwinger:1951nm}. 
In this way, the energy of the electromagnetic field is transferred to relativistic particles ($e^+e^-$) which thermalize, reheating   the universe. 
We refer to this mechanism as ``Schwinger reheating.''
If the photons are those described by the Standard Model (SM), however, it does not seem possible to reheat
to a temperature above the electron mass, which is too low for Big Bang Nucleosynthesis (BBN). 
A way to circumvent the issue is to couple, instead, the relaxion to a massless dark photon, which in turn has a small kinetic mixing with the SM photon. We show, in the second part of the paper, that the latter scenario leads to successful reheating and relaxation of the EW scale. 

The available parameter space for the above scenario, a-priori rather large, is reduced by several
 theoretical constraints that force relations among the different parameters, and by phenomenological bounds. These include 
 the validity of the effective theory, the requirement for a relaxation at the correct scale, the suppression of cosmological perturbations,  and various limits from colliders, astrophysics, cosmology and 5th-force experiments. They are discussed in Sec.~\ref{sec:constraints} and the results are presented in Fig.~\ref{Fig:constraint}.   We find that  a cutoff scale close to the Planck scale may be achieved in this relaxed inflation scenario.  

\section{Axion inflation and photon production} \label{sec:axioninfl}

In this section we review some aspects of axion inflation that are relevant to our framework. 
The inflaton will also play the role of the relaxion in the next section,
 but for now we are only interested in the dissipation mechanism due to particle production.  
We couple Abelian gauge fields to the inflaton, whose time evolution leads to the non-perturbative production of gauge field quanta.
This production has two important effects: (1) it backreacts on the inflaton and slows it down, (2) it provides a mechanism to reheat the universe at the end of inflation. Once a large number of coherent photons are produced, the reheating process follows through the  
production of $e^+e^-$ pairs via the Schwinger mechanism,
and the subsequent thermalization of the system. After that happens,
 it is important to take into account thermal effects in the gauge-boson production. 
We discuss Schwinger and thermal effects in Section \ref{sec:reheating} and in Appendix \ref{sec:thermal}.
Below, we summarize the main aspects of the gauge-field production at zero temperature. The interested reader can find more details in Refs.~\cite{Anber:2009ua, Barnaby:2011vw, Linde:2012bt, Pajer:2013fsa}.

We consider a pseudo-scalar inflaton, $\phi$,  coupled to an Abelian gauge field, in a Friedmann Robertson Walker (FRW) metric, 
\be
ds^2 \equiv g_{\mu\nu} dx^\mu dx^\nu = - dt^2 + a^2(t) d\vec x^2 = -a^2(\tau) (d\tau^2 - d\vec x^2) \, ,
\ee
with $t$ the cosmic time and $\tau$ the conformal time.
The Lagrangian reads
\be \label{Lagaxphot}
\mathcal{L} = - \frac{1}{2} \partial_\mu \phi \partial^\mu \phi -\frac{1}{4} F_{\mu\nu}F^{\mu\nu} - c_\gamma \frac{\phi}{4f} F_{\mu\nu} \tilde F^{\mu\nu} - V(\phi)  \, ,
\ee
where $F_{\mu\nu} = \partial_\mu A_\nu - \partial_\nu A_\mu$, $\tilde F_{\mu\nu} = \frac{1}{2} \epsilon_{\mu\nu\sigma\rho} F^{\sigma\rho}$, with $A_\mu$ the gauge field, and $\epsilon^{0123} = \frac{1}{\sqrt{-g}}$. The potential $V(\phi)$ will be specified in the next section.
The equation of motion for $\phi$ is given by
\be \label{phiEOM0}
\ddot \phi + 3 H \dot \phi + \frac{\partial V(\phi)}{\partial \phi} = \frac{c_\gamma}{f} \langle \vec E \cdot \vec B \rangle \, ,
\ee
where the dot denotes a derivative with respect to cosmic time $t$ and the mean field approximation is used for $\vec E \cdot \vec B$. The inflaton $\phi$ is assumed to dominate the energy density, with $\dot{\phi}^2 \ll V(\phi)$, so the Hubble
parameter is given by
\be
H \simeq \frac{\sqrt{V(\phi)}}{\sqrt{3} \Mp} \, ,
\ee
where $ M_{\rm Pl} $ is the reduced Planck mass. 

The equations of motion for the gauge field are more conveniently written using the
conformal time $\tau\equiv \int^t dt^\prime / a(t')$, which during inflation is $\tau \simeq - (a H)^{-1}$. Note that $\tau < 0$.  Choosing the Coulomb gauge  $\vec\nabla\cdot\vec A=0$, 
we have $A_0=0$ and \footnote{Neglecting $(\partial_i\phi)$ and any additional source terms, the temporal and the Coulomb gauges are equivalent, see {\it e.g.} Ref.~\cite{Adshead:2015pva}. }
\begin{eqnarray}
\label{AvecEOM}
\left(\frac{\partial^{2}}{\partial \tau^{2}}-\nabla^{2}-c_\gamma\,\frac{\phi'}{f}\,\vec\nabla\times \right)\vec A=0\,,&
\end{eqnarray} 
where a prime denotes a derivative with respect to $\tau$.
We promote the classical field $\vec A(\tau,\vec x)$ to an operator $\vec{ \hat{A}}\left(\tau,\,\vec{x}\right)$ and decompose $\vec{\hat A}$ into annihilation and creation operators 
\begin{eqnarray}\label{Afourier}
\vec{\hat A} =\sum_{\lambda=\pm}\int \frac{d^3k}{\left(2\pi \right)^{3/2}}\left[\vec\epsilon_\lambda(\vec k)\,A_\lambda^{\vec k}(\tau)\,a_\lambda^{\vec k}\,e^{i\vec k\cdot\vec x}+{\mathrm {h.c.}}\right],
\end{eqnarray}
where the helicity vectors $\vec\epsilon_\pm$ are such that $\vec k\cdot \vec \epsilon_\pm=0$ and $\vec k\times \vec \epsilon_\pm=\mp i|\vec k|\vec\epsilon_\pm$. Then, $A_\pm$ must satisfy the equation
\begin{equation}
\label{AEOM}
\frac{\partial^{2}A_\pm^{\vec k}(\tau)}{\partial\tau^{2}}+\left[k^{2}\pm 2\,k\,\frac{\xi}{\tau} \right]A_\pm^{\vec k}(\tau)=0\mbox{ ,}
\end{equation} 
where we have defined 
\begin{equation} \label{xidef}
\xi\equiv c_\gamma\frac{\dot\phi}{2\,f\,H}\,\,.
\end{equation}
The parameter $\xi$ is convenient because it stays almost constant when the term $\langle \vec E \cdot \vec B \rangle$ is the dominant dissipative force in the inflaton dynamics \cite{Anber:2009ua}. 
To set our conventions, we will assume  $\phi$ rolls from positive to negative [i.e. $V'(\phi) > 0$], so $\dot\phi < 0$ and $\xi < 0$. Furthermore, $\tau < 0$ by definition,
and we take $c_\gamma > 0$. Eq.~\eqref{AEOM} implies that low-momentum (long wavelength) modes of the $A_-$ polarization, satisfying
\be \label{tacondition}
\Omega^2 \equiv  2\,k\,\frac{\xi}{\tau} - k^{2}  >  0\, ,
\ee
develop a tachyonic instability and grow exponentially.
This condition can be rewritten as
\be \label{comoving}
k^{-1} > \frac{1}{2 |\xi|} \, (aH)^{-1} \, .
\ee
Here $k^{-1}$ is the comoving wavelength of the mode $A_-$, while $(aH)^{-1}$ is the comoving horizon, which shrinks 
during inflation.
We see that, as inflation proceeds, modes with shorter and shorter wavelength become tachyonic. Since typically  $|\xi|\lesssim {\cal O}(10)$, the comoving wavelength of the exponentially enhanced modes has a typical size comparable to the comoving
horizon. 
Note that only one polarization of the photon experiences exponential enhancement, a consequence of parity violation in the system. 
The signatures of parity violation in the CMB power spectrum have been discussed in Ref. \cite{Sorbo:2011rz}.

Eq.~(\ref{AEOM}) can be solved analytically. However, it is more illuminating to use an approximate solution, which can be derived 
from the WKB approximation,
\be \label{WKBsol}
A_-^{\vec k}(\tau) \simeq \frac{1}{\sqrt{2 \Omega(k,\tau)}} e^{\int^\tau d\tau' \Omega(k,\tau')} \, ,
\ee
valid as long as $\left\vert {\Omega'} /{\Omega^2} \right\vert \ll 1$.
The WKB solution for the tachyonic modes holds in the range $(8 |\xi|)^{-1} < |k \tau| < 2 |\xi|$, where it can be written as
\begin{equation}
\label{approx}
A_-^{\vec k}(\tau)\simeq 
\frac{1}{\sqrt{2\,k}}\left(\frac{ -k \tau}{2\,|\xi|}\right)^{1/4}e^{\pi\,|\xi|-2\,\sqrt{- 2|\xi| \,k \, \tau}} \, , \qquad |\xi| > 1 \, ,
\end{equation}
and the exponential enhancement is explicit. The modes $A_+$ are not enhanced and we ignore them in what follows. 

With the explicit solutions to Eq.~\eqref{AEOM}, one can compute
\begin{eqnarray}\label{edotb0}
\langle  \vec E\cdot \vec B \rangle=-\frac{1}{4\pi^2 \, a^4}\int dk k^3 \frac{\partial}{\partial\tau}\left(\left| A_+^{\vec k}\right|^2
-\left| A_-^{\vec k}\right|^2\right),
\end{eqnarray} 
and the photon energy density
\be \label{E2B2}
\rho_\gamma = \frac{1}{2} \langle \vec E^2 + \vec B^2 \rangle = \frac{1}{4\pi^2 \, a^4}\int dk k^2 \left( \left\vert \frac{\partial}{\partial \tau} A_-^{\vec k}  \right\vert^2 + k^2 \left\vert A_-^{\vec k} \right\vert^2 \right) \, .
\ee
In the last expression, we took $A_+\simeq 0$. 

Using Eq.~\eqref{approx}, one finds~\cite{Anber:2009ua}
\begin{align}
 \langle  \vec E\cdot \vec B \rangle &\simeq \frac{7!}{2^{21}\pi^2} \frac{H^4}{|\xi|^4} e^{2\pi |\xi|}
  \simeq 2.4\times 10^{-4} \frac{H^4}{|\xi|^4} e^{2\pi |\xi|}  , \label{edotb}
 \\
 \rho_\gamma  &\simeq \frac{6!}{2^{19}\pi^2} \frac{H^4}{|\xi|^3} e^{2\pi |\xi|}
\simeq 1.4\times10^{-4}  \frac{H^4}{|\xi|^3} e^{2\pi |\xi|}  . \label{rhog}
\end{align}
Incidentally, one can show that $\langle \vec E^2\rangle \simeq \frac{8}{7}|\xi|^2 \langle\vec B^2 \rangle$, and therefore $\rho_\gamma$ is dominated by the electric field contribution. 

The evolution of the inflaton $\phi$ is dictated by the equation of motion \eqref{phiEOM0}, with the $\ddot\phi$ term typically negligible.
In Ref.~\cite{Anber:2009ua}, the authors considered the regime where the term $\frac{c_\gamma}{f} \langle \vec E \cdot \vec B \rangle$
balances the slope $V'$, meaning that the dissipation mechanism that ensures slow roll is due to gauge-boson production. 
In such a case, the backreaction of the gauge quanta on the inflaton produces perturbations that are too large, and excluded by CMB observations. On the other hand, in Ref.~\cite{Barnaby:2011vw}, the authors considered the regime in which the term
$\frac{c_\gamma}{f} \langle \vec E \cdot \vec B \rangle$ is negligible for most of the observable e-folds and slow-roll is solely due  to Hubble friction. They showed that even in this case the photon production can leave imprints on the CMB that can be measured. 

In the scenario we investigate in this paper, inflation proceeds in the following steps:
\begin{enumerate}
\item Initially, the photon production is negligible, $|\xi| \ll 1$, and $\phi$ slow-rolls because of a nearly
flat potential. In this regime, $|\ddot{\phi}|\ll H |\dot{\phi}| $ and the equation of motion is given by
\be \label{EOMreg1}
3 H \dot \phi + \frac{\partial V(\phi)}{\partial \phi} \simeq 0 \, ,
\ee
hence
\be \label{phidotreg1}
\dot\phi \simeq - \frac{V'(\phi)}{3 H} \, .
\ee
Note that $|\dot \phi|$ increases slowly since $V'(\phi)$ is roughly constant and $H$ decreases as $\phi$ rolls down its potential.
\item Eventually, $|\dot\phi|$ increases to the point where $|\xi|$ grows large enough for the backreaction of the photons to become 
important in Eq.~\eqref{phiEOM0}. This is when we enter the second regime described by the equation of motion
\be \label{EOMreg2}
\frac{\partial V(\phi)}{\partial \phi} \simeq \frac{c_\gamma}{f} \langle \vec E \cdot \vec B \rangle \, ,
\ee
with $\langle \vec E \cdot \vec B \rangle$ given by Eq.~\eqref{edotb}. The approximate solution is
\begin{align}
\xi\simeq -\frac{1}{2\pi }\ln \left[\frac{V'(\phi)f}{{2.4 \times 10^{-4}}c_\gamma H^4}\right], \label{xiconst}
\end{align}
where we have neglected a factor of $\xi^4$ inside the logarithm. We see that $\xi$ is roughly constant in this regime, 
and we have
\be \label{phidotreg2}
\dot\phi  \simeq -\frac{fH}{\pi c_\gamma}\ln \left[\frac{V'(\phi)f}{2.4 \times 10^{-4}c_\gamma H^4}\right]  .  
\ee
Unlike the previous regime, $|\dot\phi|$ now decreases with decreasing $H$.  
The produced photons have an energy density that remains roughly constant
\footnote{We keep track the coefficient of $4/7$ for the later discussion in Sec.~\ref{sec:CMB}. Although we use the WKB approximation here, we have checked that the full solution based on the Coulomb function reproduces this coefficient with only 15\% discrepancy.}
\be
\rho_\gamma \simeq \frac{4|\xi|}{7} \langle \vec E \cdot \vec B \rangle \simeq \frac{4|\xi|}{7}\frac{f V'}{c_\gamma}  \, .
\ee
Here we have used Eqs.~\eqref{edotb}, \eqref{rhog} and  \eqref{EOMreg2}. When the potential of $\phi$ drops below the value
$V(\phi) \sim \rho_\gamma$, the photon energy density becomes dominant, and we exit inflation.
\end{enumerate}

The problem with this scenario is that the produced photons have extremely long wavelength and do not thermalize via perturbative scattering processes to reheat the universe. 
From Eq.~\eqref{comoving} it follows that their typical physical momentum, $q_\gamma = \frac{k}{a}$, is
\be \label{physicalq}
q_\gamma < |\xi| H \, .
\ee
As we describe in more detail in later section, the relaxation mechanism requires values of $H \ll$ MeV close to the end
of inflation, which in turn implies $q_\gamma \ll$ MeV.
At the same time these photons have a high occupation number in the Hubble volume $\rho_\gamma /q_\gamma H^3 \gg 1$
due to the large exponential $e^{2 \pi |\xi|}$
in Eq.~\eqref{rhog}. 
This system is best described classically as an electromagnetic field.
One can show that the photons add up coherently to form a constant electric field within the horizon with magnitude 
$|\vec{E}|\sim \sqrt{\rho_\gamma}$ (see Appendix~\ref{app:electricfield} for further discussion). This electric field grows strong enough to allow for vacuum $e^+e^-$ production
via the Schwinger mechanism. This changes dramatically the picture in the second regime described above. We discuss it
in detail in Section \ref{sec:reheating}.

So far, we have described the generalities of $\phi$ playing the role of the inflaton. Our main purpose is to use this inflaton to relax the electroweak scale and, to do so, we need to add the relaxion ingredients, that come next.  In 
the rest of the paper, we explain in detail the whole cosmological evolution of the relaxion/inflaton field.


\section{A  relaxed inflation model} \label{sec:model}

The first model we consider consists of an axion field on a very flat potential, $V_{\rm roll}$, and coupled to SM photons. This pseudoscalar dominates the energy density of the universe during inflation and acts both as the inflaton and as the field that scans the Higgs mass. Additionally, there is a periodic potential $V_{\rm wig}$ that plays a crucial role in setting the VEV of the Higgs after reheating. The effective Lagrangian for our model is given by 
\be
\mathcal{L}  =  - \frac{1}{2} \partial_\mu \phi \partial^\mu \phi  -\frac{1}{4} F_{\mu\nu}F^{\mu\nu} - c_\gamma \frac{\phi}{4f} F_{\mu\nu} \tilde F^{\mu\nu} - V(\HD, \phi)  \, , \label{eq:Lagrangian} \\
\ee
with,
\beqn
V(\HD , \phi) &  =  & \mu^2(\phi) \HD^\dagger \HD + \lambda (\HD^\dagger \HD)^2 + V_{\rm roll} (\phi) + V_{\rm wig}(\phi) +V_0 \, , \\
V_{\rm roll} (\phi) & = &  m \Lambda^2 \phi + \frac{1}{2} m^2 \phi^2 + \frac{1}{6} \frac{m^3}{\Lambda^2} \phi^3 +  \cdots \, , \label{Vrolldef} \\
V_{\rm wig}(\phi) & = & \Lambda_{\rm wig}^4 \cos \frac{\phi}{f} \, . \label{Vwigdef}
\eeqn
Here, $\HD$ is the SM Higgs doublet, $\phi$ is the relaxion/inflaton field, and
\be \label{mu2def}
\mu^2(\phi)  =  g_h m \phi - \Lambda^2 \, , 
\ee
is the $\phi$-dependent squared mass parameter of the Higgs potential. The Higgs bare mass $\Lambda$ is the cutoff of the effective Lagrangian, $g_h$ is a dimensionless parameter of order one, 
and $m \ll \Lambda$. We comment on the parameter $\Lambda_{\rm wig}$ at the end of this section.
  We omit terms with $W^\pm$ and $Z$ for simplicity. In particular, there are $(\phi/f) Z_{\mu\nu}\tilde{F}^{\mu\nu}$ and $(\phi/f) Z_{\mu\nu}\tilde{Z}^{\mu\nu}$ terms that are gauge invariant but 
 the photon production is not affected by them because their effect only appears from dimension 8 operators suppressed by $m_Z^2f^2$ after integrating out $Z$. Similarly, $W$ terms do not affect the photon production.

The relaxion potential here is the same as in Ref.~\cite{Graham:2015cka} and our model inherits some of the properties of that scenario. These include a trans-Planckian field range for  $\phi$, a nearly-flat potential $V_{\rm roll} (\phi)$, and the periodic (``wiggle") potential $V_{\rm wig}(\phi) $. On the other hand, there are sharp differences that lead to stark contrast with the original proposal. First, in our case the relaxion is the inflaton itself, which allows the energy density of the universe to be of the same order as $\sim\Lambda^4$. 
Second, since the dynamics of inflation and relaxation end at  almost the same time, the classical rolling is automatically a good description when the electroweak scale is settled.  
Finally, the  relaxion stops after the end of inflation, and therefore we require a form of dissipation distinct from the Hubble friction. 
The coupling to photons provides this extra source of dissipation, and offers a novel opportunity for reheating. 

The smallness of $m$ is justified by the
fact that as $m \to 0$, the model possesses the discrete shift symmetry $\phi \to \phi + 2\pi n  f$. 
A potential of this kind was first used by Abbott~\cite{Abbott:1984qf} in an attempt to explain dynamically the smallness of the cosmological constant. Here, following~\cite{Graham:2015cka}, we use it for the EW scale instead.  
As written, the model poses some
theoretical issues \cite{Gupta:2015uea} that can be circumvented with a clockwork axion model \cite{Choi:2015fiu, Kaplan:2015fuy, Giudice:2016yja}, which we present in Appendix~\ref{app:clock}, where we also show how to map its parameters to the ones
used in this section and in the rest of the paper. 

For the given coordinate, a special point in field space is
\be
\phi_0 \equiv \frac{\Lambda^2}{g_h m} \, ,
\ee
where $\mu^2(\phi_0) = 0$. It separates the unbroken EW phase, $\phi > \phi_0$, from the broken phase $\phi < \phi_0$. 
For field values of order $\phi_0$, the small $m$ expansion in Eq.~\eqref{Vrolldef} is not well defined, as each term is of order $\sim \Lambda^4$ and generically order one corrections are expected.  
In what follows, we will only keep the term linear in $m$ and $\phi$, a choice that is justified only once we consider a UV completion of this model, such as the clockwork axion discussed in Appendix~\ref{app:clock}. 

As most
of the interesting dynamics happens near $\phi_0$, it is convenient to expand the potential around this point. We define\footnote{We
stress that $\dphi$ is still a classical field, not a quantum fluctuation.}
\be \label{dphidef}
\phi = \phi_0 + \dphi \, , \qquad  |\dphi | \ll \phi_0 \, .
\ee
We keep only the $\phi$ linear term in the potential \eqref{Vrolldef} and expand around $\phi_0$. 
The potential then reads
\be \label{Vexpand}
V(h, \dphi)  =  \frac{1}{2} \mu^2(\dphi) \ h^2 + \frac{1}{4} \lambda h^4  + m \Lambda^2 (\phi_0+ \dphi) +  \Lambda_{\rm wig}^4 \cos \frac{\phi_0 + \dphi}{f} + V_0 \, ,
\ee
where $h$ is the radial mode of $\HD$ and  $\mu^2(\dphi) = g_h m  \dphi$.
We choose $V_0$ such that the cosmological constant has the observed value $V_{\rm cc}^{\rm obs}\sim {\rm meV}^4$ once $h$ and $\phi$ settle to their VEVs: 
\beqn
\langle h \rangle & = & v = \sqrt{- \frac{\mu^2(\dphi_{\rm EW})}{\lambda}}\simeq \frac{m_W} {\sqrt{\lambda}} \, \\  
\langle \dphi \rangle & = & \dphi_{\rm EW}\simeq - \frac{m_W^2}{g_h m }  \, .
\eeqn
We have then
\be \label{V0def}
V_0 = -\frac{\Lambda^4}{g_h}+ \frac{m_W^2 \Lambda^2}{g_h} + \frac{m_W^4}{4 \lambda}+V_{\rm cc}^{\rm obs}  \, . 
\ee 
The contribution of 
$V_{\rm cc}^{\rm obs}$ to $V_0$ is a lot smaller compared to the other two terms. In what follows we take $V_{\rm cc}^{\rm obs}$
to be effectively zero.

The parameter $\Lambda_{\rm wig}$ can be written generically as
\be \label{Lam0}
\Lambda_{\rm wig}^4 \sim (yv)^n M^{4-n} \, ,
\ee
with $n>0$ and $M$ some fixed mass scale. The fact that $\Lambda_{\rm wig}^4$ depends on the Higgs VEV, $v$, is crucial: as $v$ grows, the amplitude of the wiggles becomes larger and larger up to the point where they stop the rolling of $\dphi$. This must happen when
$v$ attains the observed value of 246 GeV. 
The case of the QCD axion corresponds to $n = 1$, $y \sim y_u$ (the lightest quark yukawa), and $M \sim f_\pi$
(the pion decay constant)\footnote{In this case, $\Lambda_{\rm wig}^4 \sim M^3 (m\dphi)^{1/2}$, there is 
a singular term in the first derivative of the potential, $\partial \Lambda_{\rm wig}^4/\partial \dphi$, at $\dphi = 0$.
The singularity is evaded thanks to the 
quark condensate, $\langle \bar q_L q_R \rangle$, which provides a tadpole for the Higgs potential and results in a small, but non zero VEV even for $\mu^2 > 0$.}.
This case is excluded \cite{Graham:2015cka} because it results in $\theta_{\rm QCD} \sim 1$ and thus is 
plagued by the strong CP problem.
In the $n=2$ case, the sector responsible for generating $V_{\rm wig}$ does not break the electroweak symmetry, we 
have a two-loop wiggle-potential \cite{Espinosa:2015eda, Gupta:2015uea} also in the unbroken electroweak phase,
and the relaxation mechanism works, provided that $M <  v$.


\section{Dynamics} \label{sec:inflation} 

In this section, we discuss the cosmological evolution of the fields $\dphi$, $h$, and $A_\mu$. 
For the purpose of our study, we can treat $\dphi$ as a homogeneous classical field, but we must treat $h$ and $A_\mu$ as quantum fields.
The equations of motion are 
\beqn 
&& \delta\ddot\phi + 3 H \delta\dot\phi + \frac{\partial V(h,\dphi)}{\partial \dphi} = \frac{c_\gamma}{f} \langle \vec E \cdot \vec B \rangle \, , \label{phiEOM} \\
&& \ddot h + 3 H \dot h + \frac{\nabla^2}{a^2} h + \frac{\partial V(h,\dphi)}{\partial h} = 0 \, , \label{hEOM} \\
&& \frac{\partial^{2}A_\pm^{\vec k}(\tau)}{\partial\tau^{2}}+\left[k^{2}\pm 2\,k\,\frac{\xi}{\tau} \right]A_\pm^{\vec k}(\tau)=0\mbox{ ,}
\eeqn
Here, an overdot denotes a derivative with respect to cosmic time $t$, and $H \equiv \frac{\dot a}{a}$ is the Hubble parameter, with $a$ the scale factor. Since inflation is driven by $\dphi$ and the energy density of the universe is dominated by $V(h, \dphi)$, the Friedmann equation yields
\be \label{Hubble}
H(h, \dphi) \simeq \frac{\sqrt{V(h,\dphi)}}{\sqrt{3} \Mp} \, .
\ee

\begin{figure} [!t]
\begin{center}
 \includegraphics[scale=.32]{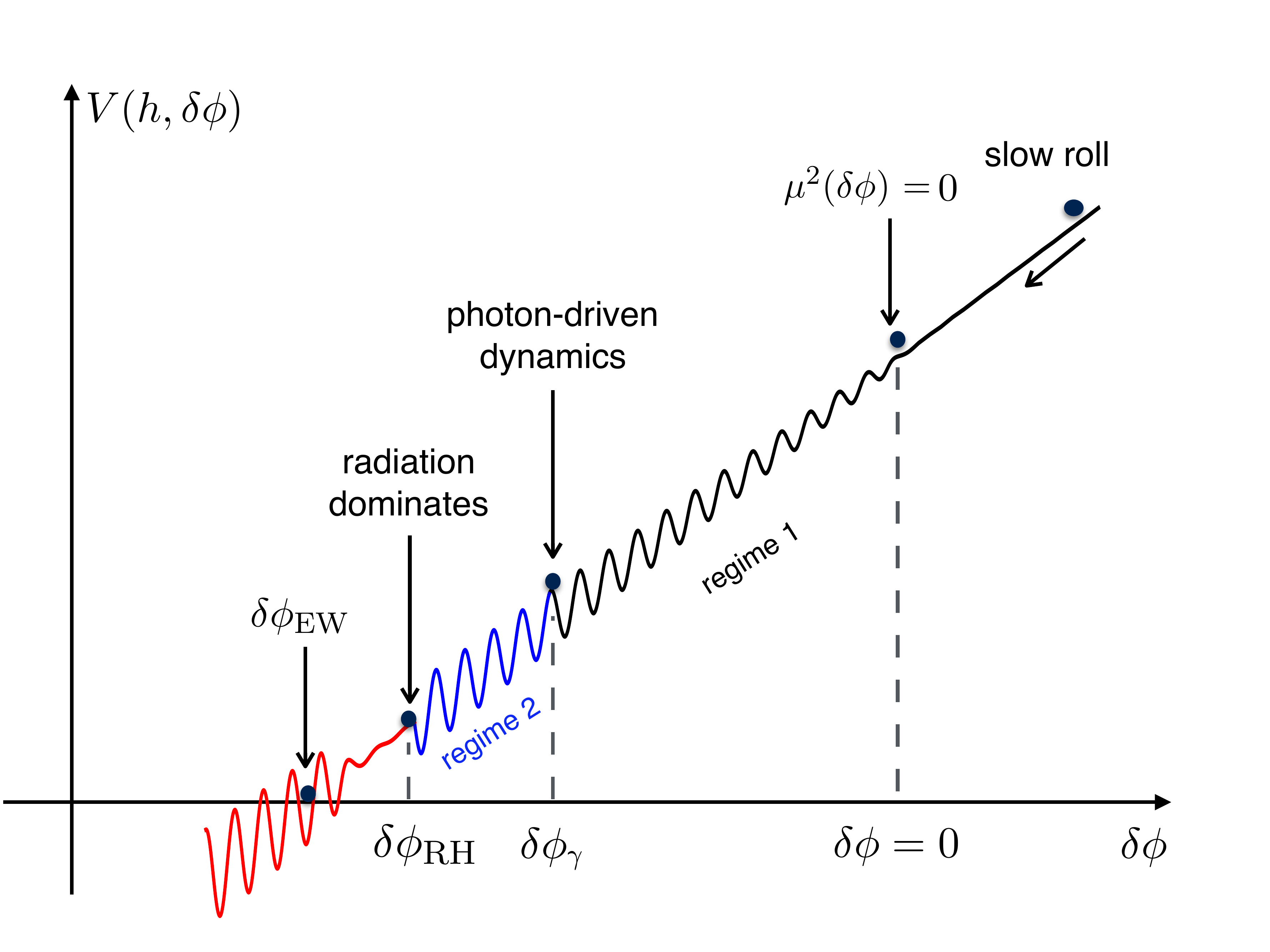}
 \caption{Sketch of the different stages in our relaxation mechanism. The first ({\bf black}) is the standard slow-roll regime, as described in subsection \ref{regime1}. In the second ({\bf blue}), the photons are responsible for the dissipation in the inflaton/relaxion dynamics, which is discussed in subsection \ref{regime2}. Finally, the last stage of relaxation occurs after reheating ({\bf red}), with the photons still providing dissipation and allowing the relaxion to get trapped in the wiggle potential (see subsection \ref{exit}).} \label{Fig:cartoon}
 \end{center}
\end{figure}

The qualitative overview of the dynamics is similar to that described in Sec.~\ref{sec:axioninfl} and is
illustrated in the cartoon of Fig.~\ref{Fig:cartoon}.
The inflaton field starts rolling from a point $\dphi_i > 0$, in the unbroken electroweak phase. In the first regime, the slow roll is 
due to the smallness of the slope $m$ and the photon production is negligible. The motion is described by $\delta\dot\phi \simeq - {V'(\phi)}/{3 H}$, with a speed $|\delta\dot\phi|$ that increases as the field rolls. We stay in this regime for a very large number of efolds
($N > 10^{30}$), all the way into the broken EW phase ($\dphi <0$). Eventually, the parameter $|\xi|= |c_\gamma\delta\dot\phi/2fH|$ grows
larger than one and we smoothly switch into the second regime, which is dominated by photon production and described by
$V'\simeq c_\gamma \langle E\cdot B\rangle/f$. We enter this regime when $\dphi$ is already very close to the end of its run, 
$\dphi_{\rm EW}$, and we remain there only for the last $\cal O$(20) e-folds. When the inflaton potential reaches $\sim |\xi|\Lambda_{\rm wig}^4/c_\gamma$, with $\xi$ roughly constant,
the energy density of the produced photons becomes dominant and we exit inflation. 
What follows is a period of radiation domination in which the relaxion keeps slowing down due to photon dissipation,
until it stops on the growing wiggles to set the observed electroweak scale.

We now give some quantitative details of each stage in this simplified picture.
In this section we neglect thermal effects, with the aim of keeping the discussion clearer.
As we will see, these effects have significant implications which 
 require a careful treatment, that we 
postpone to Section~\ref{sec:reheating}.


\subsection{Conditions on the slope $m$}

At the end of the rolling, the relaxion must stop on the wiggles. This implies two conditions on the parameters of the model:
\begin{enumerate}
\item At least one wiggle between $\dphi = 0$ and $\dphi_{\rm EW}$ must exist,
\be \label{mweak}
m < \frac{m_W^2}{f} \, .
\ee
\item Assuming significant dissipation, the inflaton must halt when the condition $|V'_{\rm roll}(\dphi)| \simeq |V'_{\rm wig}(\dphi)|$ is satisfied. Consequently, the relation
\be 
m \Lambda^2  \sim  \frac{\Lambda_{\rm wig}^4}{f} \,  \label{mwig} 
\ee
is implied.
As we mention below Eq.~\eqref{Lam0}, $\Lambda_{\rm wig}$ can never exceed $ m_W$, thus 
\be
\label{mbound}
m  <  \frac{m_W^4}{\Lambda^2 \, f}\, .
\ee
For $\Lambda \gtrsim \, m_W$, this bound is stronger than \eqref{mweak}.  
\end{enumerate}


\subsection{Regime 1: slow-roll on a gentle slope} \label{regime1}

We assume that the rolling starts from $\dphi_i > 0$. In this first regime $\delta\dot\phi$ is small, $|\xi|  \ll 1$, and we can ignore
the photon production, dropping the term $\frac{c_\gamma}{f} \langle \vec E \cdot \vec B \rangle$ in Eq.~\eqref{phiEOM}. 
Then we have
\be \label{phiEOMinf1}
\delta\ddot\phi + 3 H \delta\dot\phi + m \Lambda^2 + \frac{1}{2} g_h m \langle h^2 \rangle = 0 \, .
\ee
We can also safely drop the last term: for $\dphi > 0$, $\langle h^2 \rangle$ is zero, while for $\dphi < 0$ it never grows 
larger than $m_W^2$, which is much smaller than $\Lambda^2$.

During the slow-roll we also have $\ddot \phi \ll 3 H \dot \phi$  and therefore
\be \label{phidot0}
\delta\dot\phi \simeq -\frac{m \Lambda^2}{3 H(\dphi)} \, .
\ee
As the field rolls down the potential, $H$ decreases so $|\delta\dot\phi|$ increases.
We can introduce as usual the slow-roll parameters
\beqn
\epsilon (\dphi) & \equiv &  - \frac{\dot H}{H^2}  \simeq \frac{\Mp^2}{16 \pi}\left( \frac{V'(\dphi)}{V(\dphi)} \right)^2  \, , \\
\eta(\dphi) & \equiv & \epsilon(\dphi) - \frac{\ddot\dphi}{H\dot\dphi}  \, .
\eeqn
One slow-roll condition, $\epsilon(\dphi) < 1$, is satisfied so long as
\be \label{Vexit1}
V(\dphi) \gtrsim V_{\rm exit1} \equiv \frac{1}{4\sqrt{\pi}} \Mp V'_{\rm roll}(\delta\phi) . 
\ee
The second condition, $|\eta(\dphi)| < 1$, is also satisfied when $\epsilon(\dphi) < 1$, as we show in Appendix~\ref{sec:slowroll}.

Once $\dphi$ crosses 0, an important phenomenon happens: the Higgs field
experiences an instability, known as tachyonic or spinodal instability~\cite{Felder:2000hj, Felder:2001kt, Copeland:2002ku, GarciaBellido:2002aj}, that eventually results in the
spontaneous breaking of the EW symmetry. The instability develops fast and drives the field to the minimum of its 
mexican hat potential, while $\dphi$ has not moved much from $\dphi = 0$. From that point the dynamics of the Higgs are
well captured by the evolution of its zero mode, which oscillates around the minimum.
 Meanwhile, the minimum grows deeper, as $\dphi$ rolls to more negative values. 
 The energy density associated with Higgs oscillations grows at the expense of the 
relaxion energy density. One might wonder if in the end we store enough energy in the Higgs to allow for reheating via 
its decays into SM particles. 
The answer is negative:
the relaxion dissipates most of its energy via Hubble friction, and at the end of the run the energy density of the Higgs is still
several orders of magnitude smaller than $\Lambda_{\rm wig}^4$, insufficient to reheat above the BBN temperature.  

Another consequence of the instability at $\dphi = 0$ is that the exponential production of tachyonic modes of the Higgs
field happens at the expense of the relaxion kinetic energy, and provides another source of friction for the relaxion.
This friction is active for a very short time because, as we mentioned above, the Higgs is quickly driven to the minimum
of its potential, at which point the tachyonic production, and therefore the friction, switches off. 
The energy dissipated by the relaxion via this mechanism is absolutely negligible compared to the potential energy 
available at that point, that is $\sim m_W^2 \Lambda^2$, so it does not affect the dynamics.  

The slow-roll motion described by Eq.~\eqref{phidot0} continues into the broken EW phase, $\dphi < 0$, until 
$|\xi| = \frac{c_\gamma}{2}\frac{|\delta\dot\phi|}{H f}$ grows larger than one. 
At that point photon production becomes important. Neglecting thermal effects,
 we enter a second regime of slow-roll, where the dissipation is provided by photon production rather than Hubble friction.
We describe the associated dynamics next. 


\subsection{Regime 2: slow-roll via photon production} \label{regime2}
We switch from the first to the second regime of inflation when $|\xi|$ becomes larger than one and the increasing $|\delta\dot\phi|\simeq {V'_{\rm roll}}/{3 H}$ from Eq.~\eqref{phidot0} 
matches the $|\delta\dot\phi|$ derived assuming the photon-driven friction, Eq.~\eqref{phidotreg2},
\be
 \frac{V'_{\rm roll}}{3  H(\dphi_\gamma)} \simeq -\frac{fH(\dphi_\gamma)}{\pi c_\gamma}\ln \left[\frac{V'_{\rm roll} f}{2.4 \times 10^{-4}c_\gamma H^4}\right]\, .
\ee
This happens when the potential is
\begin{align}
 \label{Vswitch}
V_{\rm switch}& \equiv 3 \Mp^2 H^2(\dphi_\gamma)\simeq \frac{1}{2|\xi|} \, \frac{c_\gamma }{f} \Mp^2\, V'_{\rm roll}
 \sim \frac{1}{2|\xi|} \frac{c_\gamma\Mp^2}{f^2}  \Lambda_{\rm wig}^4 \, ,
\end{align}
where $\xi$, from Eq.~\eqref{xiconst}, is roughly constant. In the last equality we have used Eq.~\eqref{mwig}. 
If we compare Eqs.~\eqref{Vswitch} and \eqref{Vexit1}, we obtain
\be
\frac{V_{\rm exit1}}{V_{\rm switch}} \simeq \frac{|\xi|}{2\sqrt{\pi}}  \frac{f}{c_\gamma\Mp} < 1 \, .
\ee
The inequality is dictated by the condition \eqref{fupper} below, and implies
 that we switch to the second regime while we are still slow-rolling from the first
 ($\epsilon < 1$).
 
In the second regime, the dissipation from photon production is important and the equation of motion, Eq.~\eqref{phiEOM}, becomes
\be \label{EOM2}
V'_{\rm roll}(\dphi) \simeq \frac{c_\gamma}{f} \langle \vec E \cdot \vec B \rangle \, .
\ee
One can show that the conditions $|3 H \delta\dot\phi| \ll V'$ and $|\delta\ddot\phi| \ll V'$, leading to Eq.~\eqref{EOM2}, are satisfied
for
\be \label{fupper}
\frac{f}{c_\gamma} < \frac{\Mp}{|\xi|}  \, .
\ee 
Checking these conditions comes with some subtleties which are explained in Appendix \ref{sec:slowroll}. 

At this stage, the energy density is still dominated by the inflaton potential. From Eq.~\eqref{EOM2}, 
using Eqs.~\eqref{edotb} and \eqref{Hubble}, we find 
\be \label{xisol}
\xi \simeq - \frac{1}{2\pi} \ln \left[\frac{\xi^4}{2.4 \times10^{-4}}  \frac{f V'_{\rm roll}(\dphi)}{c_\gamma}   \, \frac{9 \Mp^4}{V^2(\dphi)} \right] \, .
\ee
The dependence on $\xi$ is largely through $\ln[V^2(\dphi)]$, and therefore $\xi$ varies little from the beginning to the exit of the second regime. 
To be more accurate, we find this value ($\equiv \xi_2 $) by using the potential \eqref{Vswitch} in Eq.~\eqref{xisol}: 
\be \label{xisw}
\xi_{\rm 2} \simeq  -\frac{1}{2\pi} \ln \left[ \frac{ 10^5 \, \xi_{\rm 2}^6}{c_\gamma^3} \, \frac{f^4}{\Lambda_{\rm wig}^4} \right] \, .
\ee
$\xi_2$ is given by the parameters of the model $c_\gamma, f, $ and $\Lambda_{\rm wig}$, and is typically ${\cal O}(20)$ in the parameter space of our interest.
%


\subsection{Inflation exit and relaxation} \label{exit}
 
From Eqs.~\eqref{EOM2} and \eqref{rhog}, we obtain that the energy density of the produced photons is
\be \label{rhogslow}
\rho_\gamma \simeq \frac{4|\xi_2|}{7}\frac{f}{c_\gamma}  V'_{\rm roll}
 \sim \frac{|\xi_2|}{c_\gamma} \Lambda_{\rm wig}^4 \, ,
\ee
where we have used Eq.~\eqref{mwig} for the last expression. 
The photon energy density remains roughly constant (up to a logarithmic variation of $|\xi|$) as the result of the approximate balance 
between the exponential production of photons and the Hubble dilution of this radiation.
Once the potential of the inflaton drops below the value 
\be \label{Vreheat}
V_{\rm RH} = \rho_\gamma   \, ,
\ee
the energy density is no longer dominated by $\phi$, we exit inflation and enter a radiation dominated universe. 
However, the photons have very low momentum and are not thermalized, hence we cannot talk about a reheat temperature yet. We address the reheating mechanism in the next section.

The motion of $\dphi$ is still described by Eq.~\eqref{EOM2}, so $|\delta\dot\phi|$ keeps decreasing as the relaxion rolls.
When the increasing amplitude of the wiggle potential reaches $\Lambda_{\rm wig}^4$ with the correct value of the EW VEV, 
the slope of the wiggles counterbalances the linear slope of $V_{\rm roll}(\dphi)$ and the relaxion stops at
\begin{align}
-\frac{\dphi_{\rm EW}}{f} \sim \frac{m_W^2\Lambda^2}{\Lambda_{\rm wig}^4}. 
\end{align}
From the end of inflation to this point, $\dphi$ has changed approximately
\begin{align}
\frac{\dphi_{\rm RH}-\dphi_{\rm EW}}{f}\sim \frac{|\xi_2|}{c_\gamma} \, .
\end{align} Given that $\delta\dot{\phi} \simeq \xi_2 H f/c_{\gamma}$, this implies that about one Hubble time has elapsed and the energy density $\rho_\gamma$ has only changed by an order-one amount.

Note this is an important difference with respect to the initial proposal of Ref. \cite{Graham:2015cka}. In that work, the relaxation of the EW scale occurs during inflation, while in ours $\phi$ settles down after the end of inflation. For this reason the friction provided by gauge-boson production is crucial in this last stage. Without it, the kinetic energy
$\frac{1}{2} \delta\dot\phi^2$ would inevitably grow larger than $\Lambda_{\rm wig}^4$ and the relaxion would overshoot the EW minimum,
causing the whole mechanism to fail.


\section{Schwinger reheating} \label{sec:reheating}

The picture described in the previous section is good for a successful dynamical relaxation of the EW scale, but fails to reheat the
universe. Each produced photon carries very little energy and the system cannot be thermalized via perturbative scattering
processes. The large occupation number of the photons implies that they form a classical electromagnetic field,
 as we explained in Section~\ref{sec:axioninfl}. 
 In order to discuss thermalization in this case, we have to take into account an important non-perturbative phenomenon: the Schwinger effect.   
We discuss it in this section and proceed to
point out a problem that arises when trying to reheat via SM photons.
In the next section we propose a resolution with a dark photon.

Quantum electrodynamics predicts that a strong electric field, $e|\vec E|\gtrsim m_e^2$, can create 
electron-positron pairs, provided that the characteristic wavelength of the photons is larger than the Compton wavelength of the electron $m_e^{-1}$.  
The virtual pairs, produced in the vacuum polarization of the photon, can be 
accelerated apart and become real asymptotic states if they can borrow enough energy from the electric field itself.
 This is known as the Schwinger effect \cite{Heisenberg:1935qt, Schwinger:1951nm}. In the presence of a constant electric field, the number of pairs produced per unit volume per unit time is \cite{Cohen:2008wz}
\begin{align}  \label{eq:schwingerrate}
\frac{\Gamma_{e^+e^-}}{V}
=\frac{e|\vec{E}|}{4\pi^3}e^{-\frac{\pi m_e^2}{e |\vec{E}|}}  \int d^2 k_\bot e^{-\frac{\pi k_\bot^2}{e |\vec{E}|}}
= \frac{(e |\vec{E}|)^2}{4\pi^3}\exp \left(\frac{-\pi m_e^2}{e |\vec{E}|}\right) \,,
\end{align}
where $\vec k$ is the electron (or positron) momentum, and $\vec k_\bot$ is the component orthogonal to $\vec E$. 

In axion inflation scenarios, like ours, one typically has very strong electric fields $e|\vec{E}| \gg m_e^2$. 
So, in a Hubble time, a large number of pairs per unit volume $\sim {(e |\vec{E}|)^2}/{4\pi^3 H}$ is produced.
In the model we consider, close to the end of the first regime,  with $1 < |\xi| \lesssim 10$, the 
electric field grows exponentially and reaches $e \langle \vec E \rangle \sim \pi m_e^2$, prompting the pair production\footnote{When pair production starts, the Higgs VEV
is almost at its final value. For this reason, it is a good approximation here to use $m_e = 0.51$ MeV for the electron
mass.}.
These electrons and positrons inherit an energy of order ${(e |\vec{E}|)}^{1/2}$, so the energy density transferred to the $e^+e^-$ pairs per unit time via the Schwinger effect is roughly ${(e |\vec{E}|)}^{5/2}$. This is a very efficient process: an order one fraction of the electric field energy density is transferred to $e^+e^-$. 
The thermalization of the produced $e^+e^-$ pairs proceeds via annihilations, $e^+ e^- \to \gamma \gamma$,
 and inverse Compton scatterings on the long-wavelength photons, $e \gamma \to e \gamma$. The rate of such processes is faster than the Hubble expansion.

 Consequently, the electrons and positrons thermalize very fast and
the temperature quickly reaches $T\sim m_e$. 

The finite temperature changes the dispersion relation of the photon, due to in-medium effects, and the tachyonic instability is suppressed, especially when the Debye mass, $m_D = e T/\sqrt{6}$,  is larger than the characteristic momentum of the instability, $m_D \gtrsim \frac{k}{a}\sim |\xi| H$.  
Accounting for these thermal effects, we arrive at different expressions for the electric and magnetic fields
 (see Appendix~\ref{sec:thermal} for details),
\begin{align}
 \frac{1}{2} \langle \vec E^2 \rangle & \simeq  \frac{1}{2 \pi^4} \frac{H^4}{m_D^4} H^4 |\xi|^9 e^{\frac{4}{\pi^2} \frac{H^4}{m_D^4} \xi^6},
 \\
 \langle \vec E \cdot \vec B \rangle  &\simeq  
  \frac{1}{2 \pi^3} \frac{H^2}{m_D^2} H^4 |\xi|^7 e^{\frac{4}{\pi^2} \frac{H^4}{m_D^4} \xi^6} \, .
\end{align}
Here $m_D \gg H$, and thus a big  suppression of order ${H^4}/{m_D^4}$ in the exponent is present when compared to the zero temperature case. 
This tells us that the intensity of the electric field  cannot go much above $|\vec{E}| \sim m_e^2/e$ because, once this threshold is crossed, the temperature reaches $T\sim m_e$  through the Schwinger effect, and thermal effect suppress the photon production. 
On top of that, since the size of the backreaction $ \langle \vec E \cdot \vec B \rangle$ is correlated with $\langle \vec E^2 \rangle$, the photon friction does not grow enough, unless $|\xi|$ reaches the very large value $(m_D/H)^{2/3}$.  
 
 Now we have two issues: (1) because of the suppressed backreaction, the relaxion does not slow down enough
and does not stop on the wiggles (its kinetic energy at the end is larger than the height of the barriers, $\frac{1}{2} \delta\dot\phi^2 > \Lambda_{\rm wig}^4$), (2) the reheat temperature would be of order $m_e$, which is below BBN temperature. 
One way to fix both problems is to introduce a dark photon, as we describe in the next section.


\section{A model with a dark photon}\label{sec:darkphoton}

We have seen that the scenario where the relaxion couples to the SM photon is not viable due to thermal effects. 
In this section we show that by coupling, instead, the relaxion to a dark photon, we can avoid those issues and successfully
achieve relaxation of the EW scale and reheating. 
We consider the following Lagrangian 
\begin{align} 
\mathcal{L} \, = & \, -\frac{1}{2} \partial_\mu \phi \partial^\mu \phi  -\frac{1}{4} F_{\mu\nu}F^{\mu\nu} -\frac{1}{4} F_{D,\mu\nu}F_D^{\mu\nu} -\frac{\kappa}{2}F_{D,\mu\nu} F^{\mu\nu} - c_{\gamma_D} \frac{\phi}{4f} F_{D,\mu\nu} \tilde F_D^{\mu\nu}  \nonumber \\
& +e A_\mu \bar \psi_e \gamma^\mu \psi_e - V(\HD, \phi)  \, , \label{Eq:Lag_dark}
\end{align}
where the index $D$ denotes the massless dark photon. Here, $\psi_e$ is the visible electron, and we assume there is no light matter
content in the dark sector besides the dark photon. The field redefinition $A_\mu \to A_\mu - \kappa A^D_\mu$ removes the kinetic
mixing and introduces a coupling of the dark photon to the visible electrons, $e \kappa A^D_\mu \bar \psi_e \gamma^\mu \psi_e$. 
Note that the coupling of the dark photon to $\phi$ distinguishes it from the visible photon.  
Since during the cosmic evolution only dark photons are produced in the time-dependent $\phi$ background, our choice of shifting only the visible photon in order to remove the mixing proves convenient.
 
The relevance of the photons being dark clarifies when describing the end of inflation and reheating. 
They are produced in the same fashion as described in the first part of the paper, and give rise to a constant dark electric field
$|\vec{E}_D|\sim \sqrt{\rho_{\gamma_D}}$. 
The equations derived in Sections \ref{sec:axioninfl} and \ref{sec:inflation} can be used for this model simply with the replacements: $c_\gamma \rightarrow c_{\gamma_D}$, $e \rightarrow \kappa e$.
In particular, because the coupling to electrons is suppressed by $\kappa$, the Schwinger
production rate is now
\begin{align}  \label{eq:schwingerrateDark}
\frac{\Gamma_{e^+e^-}}{V}
= \frac{(\kappa e |\vec{E_D}|)^2}{4\pi^3}\exp \left(\frac{-\pi m_e^2}{\kappa e |\vec{E_D}|}\right) . 
\end{align}
It becomes effective at larger values of $|\vec{E}_D|$, compared to the SM photon case, when
\be 
\kappa e |\vec{E}_D| > \pi m_e^2.
\ee 
The maximum value the dark electric field can achieve is given by $|\vec{E}_D^{\rm max}| \sim \left( \frac{|\xi_2|}{c_{\gamma_D}} \right)^{1/2}\Lambda_{\rm wig}^2$ (see Eqs.~\eqref{rhog}, \eqref{mwig} and \eqref{EOM2}) and consequently, 
\be 
\kappa e \,\gtrsim\, \frac{m_e^2}{\Lambda_{\rm wig}^2}  \left( \frac{c_{\gamma_D}}{|\xi_2|} \right)^{1/2} \, . \label{Eq:ek1}
\ee

To avoid the complication we encountered with the suppressed tachyonic production of visible photons, we wish to ensure that there is no thermal mass associated with the dark photon. 
To do so, we require the dark photon to be sufficiently weakly coupled as to stay out of thermal equilibrium.  Equivalently, the dark photon's mean free path, $\ell_{\rm m.f.p.}$,  must be larger than the Hubble radius, and therefore it cannot be refracted.  
Such a condition reads
\be \label{mfp}
\ell_{\rm m.f.p.} \equiv \frac{1}{n_e \sigma_{e\gamma_D\to e\gamma}}\sim \frac{1}{\kappa^2\alpha^2 ~T} > \frac{1}{H},
\ee 
and needs to hold until the relaxion settles down. This is satisfied as long as 
\be 
 \kappa e \,\lesssim\, \left(\frac{\Lambda_{\rm wig}}{\alpha \Mp}  \right)^{1/2} \left(\frac{|\xi_2|}{c_{\gamma_D}} \right)^{1/8}.\label{Eq:ek2} 
 \ee 
 In Eq.~\eqref{mfp}, we took the electrons to be relativistic and in thermal equilibrium at a temperature $T > m_e$, so that their number
 density $n_e$ scales as $T^3$. We considered the cross section $\sigma_{e\gamma_D\to e\gamma} \sim \frac{\kappa^2\alpha^2}{T^2}$,
 rather than $\sigma_{e\gamma_D\to e\gamma_D} \sim \frac{\kappa^4 \alpha^2}{T^2}$, since the latter is suppressed by two extra powers
 of $\kappa$. Also, we took $H\sim \frac{T^2}{\Mp}$ and used the reheating temperature $T \sim \left(|\xi_2|/c_{\gamma_D} \right)^{1/4}\Lambda_{\rm wig}$, since it changes only by an order-one amount between reheating and the end of relaxation, as explained in Sec. \ref{exit} . Note that with these choices the bound \eqref{Eq:ek2} is conservative.
 
The absence of a thermal mass for the dark photons implies that we keep producing efficiently the dark electric field as 
we enter the second regime of slow roll for the relaxion, described in Section~\ref{sec:inflation}, where the main friction force 
arises from dark photon production. We saw that in this regime the amount of energy available in the dark electric field is
 \be \label{rhogd}
 \rho_{\gamma_D} \,\sim \,\frac{|\xi_2|}{c_{\gamma_D}} \Lambda_{\rm wig}^4.
 \ee 
 The energy transfer from the dark electric field to $e^+e^-$ directly by the Schwinger effect is inefficient, unlike in the SM photon case. In a Hubble time this can be estimated as
\begin{align}
\frac{\Delta \rho_{\rm Schwinger}}{\rho_{\gamma_D}}
\sim \frac{(\kappa e |\vec{E}_D|)^{5/2}H^{-1}}{|\vec{E}_D|^2}
\sim \frac{(\kappa e)^{5/2}{\Mp}}{|\vec{E}_D|^{1/2}}
\lesssim \frac{1}{\alpha}\left(\frac{c_{\gamma_D}}{\alpha |\xi_2|}\frac{\Lambda_{\rm wig}}{\Mp}\right)^{1/4} \ll 1 \, ,
\end{align}
where the typical $e^\pm$ energy is $ (\kappa e |\vec{E}_D|)^{1/2}$, and we used Eq.~\eqref{Eq:ek2}  in the inequality. 

However, the electric field can transfer an amount $(\kappa e |\vec{E}_D|)d$ of energy to each electron, by accelerating it over a distance $d$.
 Shortly after Schwinger creation, the number density of electrons is 
 $n_e =  \frac{\Gamma_{e^+e^-}}{V} \Delta t \sim (\kappa e |\vec{E}_D|)^2 H^{-1}$. Thus, the energy density transferred can be 
 estimated as
\begin{align}
 \frac{n_e\cdot  (\kappa e |\vec{E}_D|)d}{\rho_{\gamma_D}}
 \sim (\kappa e)^3 \left(\frac{\Mp} {\Lambda_{\rm wig}} \right)^{2}\frac{d}{H^{-1}}
  \lesssim \alpha^{-3/2} \left(\frac{\Mp} {\Lambda_{\rm wig}} \right)^{1/2}\frac{d}{H^{-1}} \sim 10^{11} \frac{d}{H^{-1}} \, ,
\end{align}
where we took $H^2\sim \rho_{\gamma_D}/\Mp^2\sim \Lambda_{\rm wig}^4/\Mp^2$, with $\Lambda_{\rm wig} = 100$ GeV, and we again used Eq.~\eqref{Eq:ek2}. This is very efficient, provided that 
\begin{equation} \label{Eq:trans_kappa}
\kappa e > \left( \frac{\Lambda_{\rm wig}}{\Mp} \right)^{2/3},
\end{equation}
and implies that an order one fraction of $\rho_{\gamma_D}$ can be quickly transferred to the SM radiation,
so the reheating temperature can reach
  \be  \label{TRH}
 T_{\rm RH}\,\sim\,  \left(\frac{|\xi_2|}{c_{\gamma_D}}\right)^{1/4}\Lambda_{\rm wig}.
\ee

One can show that for values of the kinetic mixing bounded by Eqs.~\eqref{Eq:ek1} and \eqref{Eq:ek2}, the dark photons never reach
thermal equilibrium with the visible sector, after reheating of the latter, and remain cold. 
So far we have assumed a massless dark photon to maximize its production via the relaxion. However, one can give it a small mass. 
Its mass would have to be small enough in order not to suppress significantly its production, otherwise the relaxation mechanism could be spoiled.

\section{Cosmological perturbations} \label{sec:CMB}

Our model is similar to those of natural inflation, where the axion field couples to Abelian gauge bosons. The associated cosmological perturbations have been largely investigated in the literature~\cite{Anber:2009ua, Barnaby:2011vw,  Meerburg:2012id, Linde:2012bt, Pajer:2013fsa, Adshead:2015pva, Notari:2016npn, Garcia-Bellido:2016dkw}.
The coupling $\phi F\tilde F$ leads to several features, which include the generation of curvature perturbations and nongaussianities, 
the production of gravitational waves, and the formation of
primordial black holes (PBH). See Ref.~\cite{Pajer:2013fsa} for a review of these topics. 

In most models of natural inflation the Hubble scale is of order $10^{13}$ GeV, and the number of e-folds is roughly 60. 
The important difference in our model is that the potential is much shallower. At the beginning we can also have $H \sim 10^{13}$ GeV, but then inflation proceeds for more than $10^{30}$ e-folds and most of the potential energy initially stored
in the scalar field is dissipated. The energy density $V_*$ available close to the end of inflation is of order 
$\Lambda_{\rm wig}^4 < m_W^4$. 
For this reason, our model should be regarded as a low-scale inflation model. The number of observable e-folds is given 
by~\cite{Liddle:1993fq}
\be \label{Nk}
N(k) = 62 - \ln \frac{k}{a_0 H_0} - \ln \frac{10^{16} \ {\rm GeV}}{V_*^{1/4}} + \ln \frac{V_*^{1/4}}{V_{\rm end}^{1/4}} - \frac{1}{3} \ln\frac{V_{\rm end}^{1/4}}{\rho_{\rm RH}^{1/4}} \, ,
\ee
where $V_*$ is the energy density when the mode $k$ left the horizon, $V_{\rm end}$ the energy density at the end of inflation, $\rho_{\rm RH}$ the energy density at reheating, and the subscript 0 refers to today's value.
In our case we have $V_* \sim V_{\rm end} \sim \rho_{\rm RH} \sim \Lambda_{\rm wig}^4$. Taking the highest
value for $\Lambda_{\rm wig}$, that is $\Lambda_{\rm wig} \sim m_W$, we have
 $\ln \frac{10^{16} \ {\rm GeV}}{V_*^{1/4}} \simeq 32$, while the other logarithms in Eq.~\eqref{Nk} are roughly zero.
Therefore the observable number of observable e-folds in our model is $N(k) \simeq 30$.

We have two sources for curvature perturbations: one is from vacuum quantum fluctuations $\delta \varphi$ of the
inflaton, proportional to $H$, the other
is from fluctuations induced by the inverse decay of photons~\cite{Barnaby:2011vw}, $\delta A + \delta A \to \delta \varphi$.
The first one gives a power spectrum which, as we show in Appendix~\ref{app:curv}, is largely insufficient to explain the
observed perturbations:
\be
{\cal P} = \frac{H^4}{4\pi^2 \dot\phi^2} < 10^{-48} \frac{m_W}{f} \ll {\cal P}_{\rm COBE} = 2.5 \times 10^{-9} \, .
\ee
The smallness of ${\cal P}$ here is a consequence of low-scale $H$, combined with a very shallow potential.
Including the second contribution in regime 1, we have~\cite{Barnaby:2011vw}
\begin{equation} \label{PkPeloso}
P_\zeta (k) = {\cal P} \left( \frac{k}{k_0} \right)^{n_s -1} \left[ 1+ {\cal P} f_2(\xi) e^{4\pi |\xi|} \right] \simeq  {\cal P}^2 f_2(\xi) e^{4\pi |\xi|} \,  \quad (\text{regime 1})
\end{equation}
where $k_0 = 0.002 \ {\rm Mpc}^{-1}$ and $f_2(\xi) \simeq {10^{-4}}/{\xi^6} $. 
The second equality in \eqref{PkPeloso} holds for large $|\xi|$, and for the sake of the estimate we took $n_s \simeq 1$. 
As $|\xi|$ increases, the power spectrum \eqref{PkPeloso} increases exponentially. When we
enter regime 2, $\xi$ remains quasi-constant with value $\xi_2$ [see \eqref{xisw}], and the power spectrum saturates
to \cite{Anber:2009ua, Linde:2012bt}
\begin{align}
P_\zeta (k)  \simeq   \frac{1}{(2\pi \xi_2)^2}\sim 5\times 10^{-5} \gg {\cal P}_{\rm COBE} 
\quad (\rm \text{regime 2}) \label{Preg2}
\end{align}

\begin{figure} [!t]
\begin{center}
 \includegraphics[scale=.6]{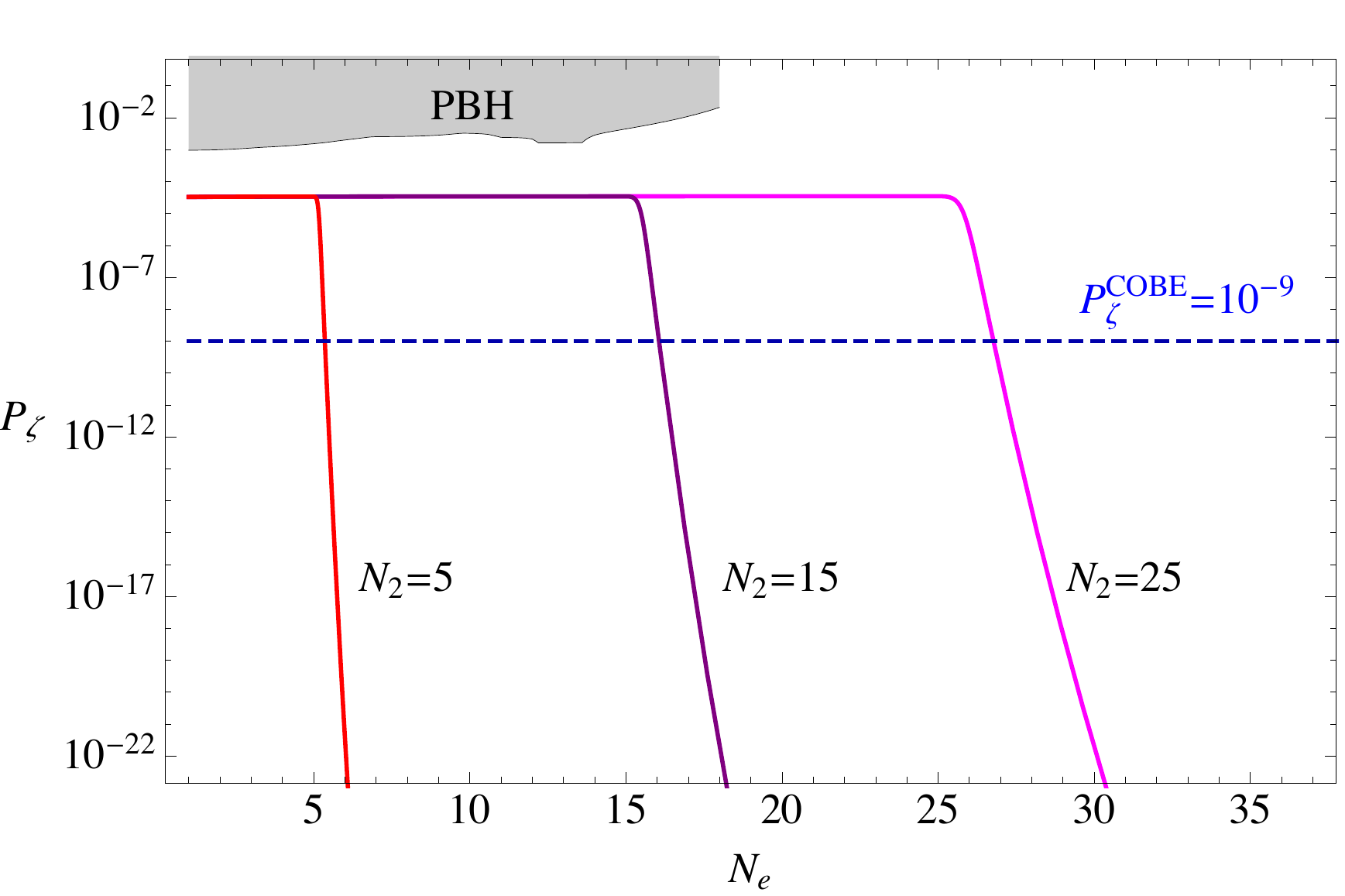}
 \caption{We show the power spectrum as a function of the number of e-folds from the end of inflation. We have fixed
 $\Lambda_{\rm wig} = 10$ GeV, and $f = 10^{16}$ GeV. The three curves  correspond to values of the 
 axion-photon coupling, $c_{\gamma_D}$, so that the number of e-folds in regime 2 [see Eq.~\eqref{N2bound}]
  is 5, 15, and 25, respectively. 
  The curves are flat in regime 2 [see Eq. \eqref{Preg2}], and fall exponentially in regime 1 [Eq. \eqref{PkPeloso}].   On the  horizontal axis, time increases
  from right to left.
  The gray region shows the bound from PBH, computed following 
  Refs.~\cite{Linde:2012bt, Garcia-Bellido:2016dkw}. Such a bound is typically very stringent for natural inflation models, but 
  here  it is easily evaded due to the significantly lower inflation scale. 
  Note that while it is numerically possible to explain the observed $P_\zeta = {\cal P}_{\rm COBE}$ around 
  $N_e = 30$, we see that the lines fall very steeply, due to the exponent in Eq.~\eqref{PkPeloso}. This indicates that
  it is difficult for the model as it stands to predict the observed curvature perturbations.
   }
  \label{Fig:Powerspectrum} 
 \end{center}
\end{figure}

This allows, in principle, to have a period around 30 e-folds from the end of inflation where we are still in regime 1, 
but with a large
 $|\xi|$ such that, thanks to the exponential in \eqref{PkPeloso}, we match the observed power spectrum for curvature
perturbations, $P_\zeta (k) \sim {\cal P}_{\rm COBE}$. In practice, it still does not mean that this model is agreement 
with CMB observations, unfortunately. The same exponential implies, as we see in Fig.~\ref{Fig:Powerspectrum},  
that $P_\zeta$ changes by many orders of magnitude within a couple of e-folds, which is in contradiction with 
CMB measurements of higher multipoles. Therefore, we need to roughly have less than 25 e-folds in regime 2
to comply with CMB bounds, the consequence being that we do not produce the observed amount of curvature 
perturbations in the model as it stands. We note that the addition of another field, like a curvaton, 
can help in matching the CMB power spectrum. 

Our current estimate does not take into account the modulation effects due to the wiggles~\cite{Flauger:2009ab}. 
Naively they are negligible, because
${\cal P}$ is so small, but a conclusive statement requires a dedicated study, beyond the scope of this paper.
We leave a more detailed study of the parameter space in relation to CMB constraints, and a possible extension of 
this model, to future study.



\section{Constraints and relevant scales} \label{sec:constraints}

\begin{figure} [!t]
\begin{center}
 \includegraphics[scale=.6]{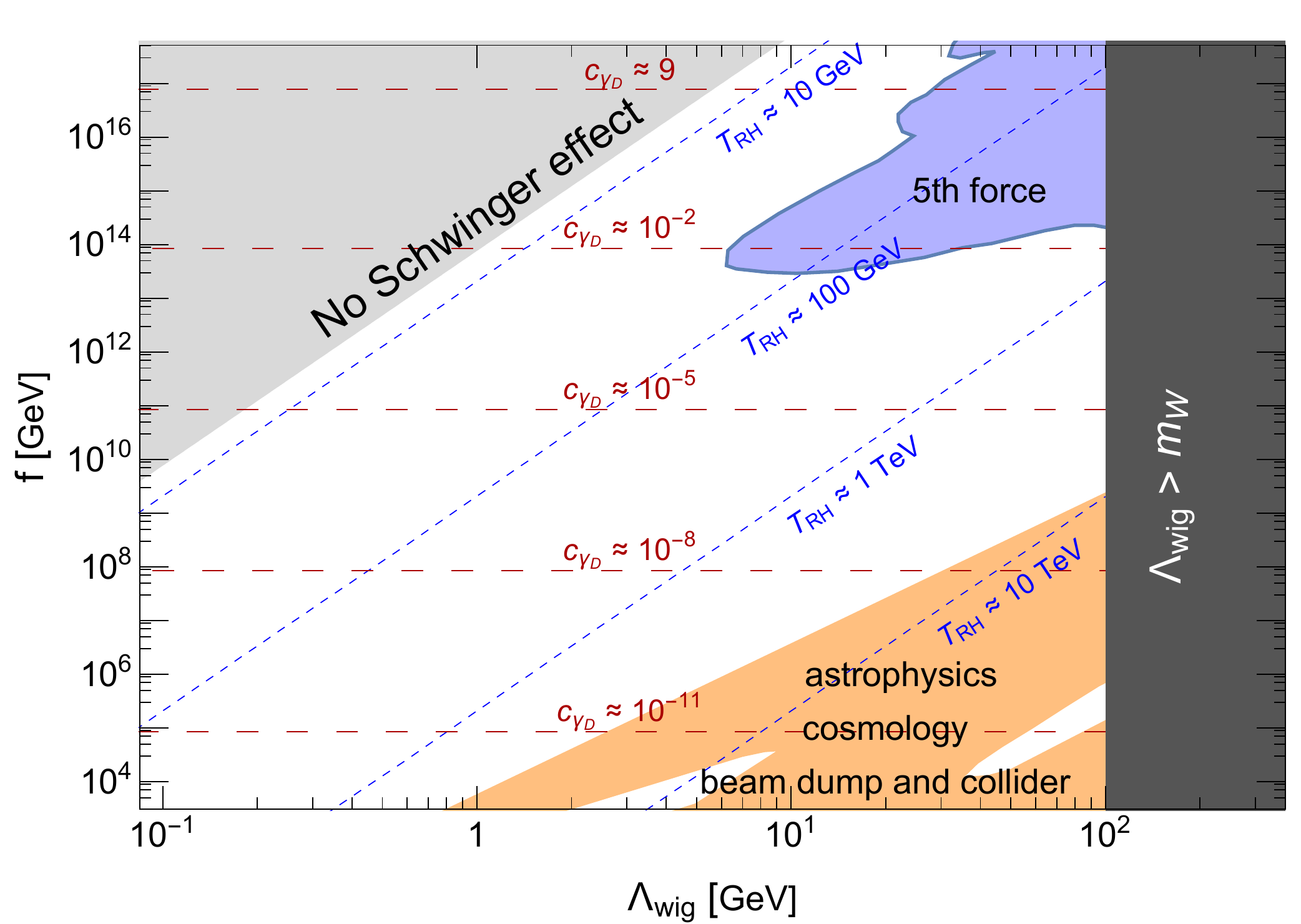}
 \caption{Summary of constraints on our model. We fix $f/{c_{\gamma_D} = 0.2\Mp}/{ |\xi_2|}$. The range of $|\xi_2|$ in this plot is from 19 to 26. The dashed red contours show the values of $c_{\gamma_D}$ that were chosen to saturate the lower bound in Eq. \eqref{fovawindow}. The dashed blue contours depict the reheating temperature given in Eq. \eqref{Eq:Trh2}. The blue region is excluded by 5th force constraints, while the orange region corresponds to a set of constraints from astrophysics, cosmology, beam dump experiments, and colliders; these are explained in detail in \cite{Flacke:2016szy}. The dark gray area corresponds to values of $\Lambda_{\rm wig}\, >\, m_W$ and is excluded as it implies an unacceptable electroweak breaking scale. The light gray region is defined by the bound Eq.~\eqref{lowerLbr}, combined with
 Eq.~\eqref{fovawindow}. In such a region there is no viable value of the mixing $\kappa$ to allow at the same time for reheating via the Schwinger
 effect and for the dark photon to avoid a thermal mass.
 Note that of the dimensionful parameters listed in \eqref{modelpar}, $\Lambda$, is fixed by Eq.~\eqref{mwig},  and $m$ does not need to be specified as long as Eq.~\eqref{mlowup} is satisfied, which is always the case.
  }
  \label{Fig:constraint} 
 \end{center}
\end{figure}

We are now in the position of summarizing the constraints on the model with the dark photon. 
A summary plot is given in Figure \ref{Fig:constraint}. The independent parameters in our construction are
\be \label{modelpar}
m,\,  \Lambda, \, g_h, \,  \Lambda_{\rm wig}, \, f, \, c_{\gamma_D},  \, \kappa.
\ee 
The first 3 parameters are related to the shallow rolling potential, $\Lambda_{\rm wig}$ and $f$ are related to the wiggle potential, $c_{\gamma_D}$ and $\kappa$ to the hidden photon coupling to $\phi$ and the visible sector respectively.
For the sake of simplicity, we take $g_h = {\cal O}(1)$.
Since $\phi$ is the Goldstone of a global symmetry spontaneously broken at the scale $f$, we must impose that the scale 
$\Lambda$, which explicitly breaks the symmetry, be smaller than $f$, 
\be \label{L<f}
\Lambda \lesssim f \, .
\ee
This implies a lower bound on $m$ 
from Eq.~\eqref{mwig}. We also require the presence of many wiggles between $\dphi = 0$ and $\dphi_{\rm EW}$, that is we impose
$|\dphi_{\rm EW}| > f$. This implies an upper bound on $m$. The two conditions together give the window
\be \label{mlowup}
\frac{\Lambda_{\rm wig}^4}{ f^3} \lesssim m < \frac{m_W^2}{f} \, ,
\ee   
with $\Lambda_{\rm wig} < m_W$, as discussed at the end of Section~\ref{sec:model}.

The combination $f/c_{\gamma_D}$ is constrained to the window 
\be \label{fovawindow}
\frac{0.1 }{|\xi_2|} \Mp \lesssim \frac{f}{c_{\gamma_D}} < \frac{1}{|\xi_2|} \Mp \, .
\ee
The upper bound comes from the requirement that we enter the photon-dominated slow-roll regime, while the lower
bound comes from asking that such a regime does not last more than the last 25 e-folds, see Fig.~\ref{Fig:Powerspectrum}.

The number of e-folds in regime 2 is
\be
N_2 = \int_{t_{\rm switch}}^{t_{\rm RH}} H dt = \int_{V_{\rm switch}}^{V_{\rm RH}} \frac{H}{\delta\dot\phi V'} dV
= \int_{V_{\rm switch}}^{V_{\rm RH}} \frac{c_{\gamma_D}}{2\xi f V'}dV  \, ,
\ee
with $V_{\rm switch}$ given by Eq.~\eqref{Vswitch} and  $V_{\rm RH}$ by Eq.~\eqref{Vreheat}. 
Requiring this to last for 25 e-folds at most, 
and treating $\xi$ as a constant, gives
\be \label{N2bound}
N_2 \simeq \frac{c_{\gamma_D}}{2\xi_2 f V'} \int_{V_{\rm switch}}^{V_{\rm RH}} dV
 \simeq 
 \frac{c^2_{\gamma_D}}{4 |\xi_2|^2 f^2} {\Mp^2} - 
\frac{2}{7}  < 25 \, ,
\ee
from which we obtain the lower bound of \eqref{fovawindow}.
Incidentally, in this window we have $V_{\rm switch}/V_{\rm RH}\sim \frac{c^2_{\gamma_D}}{f^2}\frac{\Mp^2}{ |\xi_2|^2 } \sim {\cal O}(10)$, which confirms,
following Eq.~\eqref{xisol}, that $\xi$ varies very little during this regime. We stress that since $c_{\gamma_D}$ is a free parameter, this rather narrow window leaves a significant viable parameter space, nonetheless.

The goal of the whole mechanism is to achieve a cutoff $\Lambda$ as large as possible. As $\Lambda \lesssim  f$, the cutoff
is only limited by the upper bound on $f$ from Eq.~\eqref{fovawindow}. We have seen that $|\xi|$ varies only logarithmically
in the short photon-dominated regime, and its value is typically $|\xi_2| \sim {\cal O}(20)$. To increase the allowed value of $f$ one would
like a value of $c_{\gamma_D}$ as big as possible. Large values of $c_{\gamma_D}$ can possibly be achieved in the clockwork framework, 
see Appendix~\ref{app:clock}, but for now we
restrict our attention to the case $c_{\gamma_D} < 10$. Note that once we fix $c_{\gamma_D}$ we get directly an upper bound on $f$ and 
on the cutoff $\Lambda$, independently of the other parameters of the model.

Finally, as discussed in Section~\ref{sec:darkphoton}, we need to ensure that the dark photons create $e^+e^-$ pairs, Eq.~\eqref{Eq:ek1}, while not acquiring thermal mass, Eq.~\eqref{Eq:ek2}. We also impose that the dark electric field transfers sufficient energy to the $e^+e^-$, Eq.~\eqref{Eq:trans_kappa}. 
Together,  these requirements contrain $\kappa e$ to the window,
\begin{align}
{\rm max}\left[\frac{m_e^2}{\Lambda_{\rm wig}^2} \left( \frac{c_{\gamma_D}}{|\xi_2 |} \right)^{1/2}\ , \  \left(\frac{\Lambda_{\rm wig}}{M_{\rm Pl}}\right)^{2/3} \right]
\lesssim \kappa e \lesssim \left(\frac{\Lambda_{\rm wig}}{\alpha \Mp}  \right)^{1/2} \left(\frac{|\xi_2|}{c_{\gamma_D}} \right)^{1/8}
\end{align}
Any value of $\kappa$ in this range will be equally good for reheating. 
At the same time, they yield a lower bound on $\Lambda_{\rm wig}$,
\be \label{lowerLbr}
\Lambda_{\rm wig} > \left(\alpha \Mp m_e^4 \right)^{1/5} \left(\frac{c_{\gamma_D}}{|\xi_2 |} \right)^{1/4}.
\ee 
The reheating temperature we get is (see Eq.~\eqref{rhogd})
\begin{align} \label{Eq:Trh2}
T_{\rm RH}\,\sim\, \left(\frac{|\xi_2|}{c_{\gamma_D}}\right)^{1/4} \Lambda_{\rm wig}. 
\end{align}

We provide a benchmark point to give an idea of the scales and numbers involved. 
First, we fix $f/{c_{\gamma_D} = 0.2\Mp}/{ |\xi_2|}$, $\Lambda = 0.1 f$ (which means
  $m \sim 100 {\Lambda_{\rm wig}^4}/{f^3}$), and take for instance, $\Lambda_{\rm wig} = 1$~GeV and $f= 10^{11}$~GeV. 
  Thus, we have 
\begin{align}
& g_h = {\cal O}(1)\, , \, \Lambda_{\rm wig} = 1 \ {\rm GeV}\, ,\, f =  10^{11} \ {\rm GeV}\, ,  \, \Lambda =  10^{10} \ {\rm GeV} \, , \nonumber \\
&c_{\gamma_D}\simeq10^{-5} \,  , \,
m \sim 10^{-31} \ {\rm GeV}\, ,  \,  10^{-10} \lesssim\kappa\lesssim 10^{-9} \, ,\nonumber \\
&|\xi_2| \simeq 24 \, , \, T_{RH}\sim 26 \ {\rm GeV} .  \label{benchQCD}
\end{align}
More generically, we show in Fig.~\ref{Fig:constraint} the allowed (white) region on the plane $f$ vs $\Lambda_{\rm wig}$. Note 
that for $\Lambda_{\rm wig} \sim 10$ GeV, values of $f$ very close to $\Mp$ are allowed, which in turn can accommodate 
a cutoff as high as $10^{16}$ GeV.

The relaxion mass and its mixing angle with the Higgs are given by 
\begin{align}
\label{relaxionmass}
m^2_\phi \sim  \frac{\Lambda_{\rm wig}^4}{f^2}, 
\quad \theta \sim \frac{\Lambda_{\rm wig}^4}{f m_W^3}, 
\end{align}
Here the contributions from $V_{\rm roll}$ are negligible: they are small because they break the discrete shift symmetry. 
For high values of $f$, say above $10^{10}$ GeV, 
the relaxion mass is smaller than 1~keV, and its couplings to matter, suppressed by $\theta\lesssim10^{-8}$, are tiny. 
In this range it is hard to detect it experimentally as a particle.
However, via its mixing with the Higgs, it can be the mediator of a long-range force. Experimental tests for fifth force \cite{Piazza:2010ye, Flacke:2016szy} (blue in Fig.~\ref{Fig:constraint}), provide interesting bounds for high $f$. To cover the whole region
with $f>10^{14}$ GeV their sensitivity would have to improve by a few orders of magnitude.
For $f\lesssim 10^9~{\rm GeV}$, the mass of the relaxion is above 10~keV. In this region of parameter space, the relaxion can be probed via cosmological and astrophysical processes, or in the laboratories, and there are various constraints studied in Refs.~\cite{Choi:2016luu, Flacke:2016szy} (orange in Fig.~\ref{Fig:constraint}). 

Concerning the dark photon, there are almost no experimental constraints in our scenario. This is because the dark photon
has to be massless or extremely light, $m_{\gamma_D} < 10^{-14}$~GeV, and the mixing very small, $\kappa < 10^{-8}$ 
(see {\it e.g.} \cite{Harnik:2012ni} for bounds that extend to this region of parameter space). 

CMB observables represent perhaps the most interesting arena for testing this framework. The dark photon production can
lead to the generation of nongaussianities, primordial black holes and gravitational waves, while the wiggles of the relaxion 
potential can produce measurable modulations. These features deserve a dedicated study, which is beyond the scope 
of the current work.

In this relaxed inflation scenario, we can achieve a higher cutoff than in Ref.~\cite{Graham:2015cka}.
The limiting factors in the original model were the conditions:
\begin{enumerate}
\item that the vacuum energy be dominated by the inflaton,
\item that the evolution of the relaxion be dominated by classical rolling rather than quantum fluctuations,
\item that the Hubble parameter during inflation be smaller than $\Lambda_{\rm wig}$ for the wiggles to appear.
\end{enumerate}  
In the framework presented in this paper, these three conditions are not relevant, so we can achieve a cutoff
$\Lambda \sim 10^{16}$ GeV. It is obvious why condition 1 does not apply, as in our case the relaxion is the inflaton itself.
Condition 2 is not necessary since $\dphi$ settles down when the universe is not de Sitter anymore but radiation dominated. 
Condition 3 is not necessary either, as our wiggles reappear after reheating once the temperature drops below $\Lambda_{\rm wig}$.


\section{Summary} \label{sec:summary}

We have investigated a model in which the relaxion, originally proposed in Ref.~\cite{Graham:2015cka}, is also the inflaton.
Two key ingredients of the original proposal were a very shallow slope of the potential and the presence of a periodic potential (wiggles), with
amplitude growing proportionally to the Higgs VEV. The wiggles provide the backreaction necessary to stop the motion of 
the relaxion and set the observed EW scale. 
A shallow slope suggests that the relaxion itself could be the inflaton, as it automatically satisfies the slow-roll   
conditions. 
The EW scale must be set after the end of inflation and to avoid overshooting it is necessary to introduce an additional dissipation mechanism. We have shown that this can be accomplished
by coupling the relaxion to gauge bosons. In the last stages of inflation, the  gauge-boson production becomes significant, slowing down the relaxion and allowing for a new reheating mechanism.

The reheating process is an important novelty of this work. It first starts with the production of very strong electric and magnetic fields,
which allow for vacuum electron-positron pair production via the Schwinger mechanism. The $e^+ e^-$ pairs quickly thermalize, reheating the universe.  
To achieve a sufficiently high reheat temperature, the produced gauge bosons cannot be coupled strongly to the thermal bath, as thermal effects quickly shut off the non-perturbative photon production.  Here we considered the production of dark photons which are only weakly coupled to the visible sector.  We find that this allows to reheat safely above BBN temperature, while the unsuppressed production
of dark photons provides enough dissipation for the relaxion, which slows down and settles on the correct EW minimum.
A detailed study of this reheating mechanism is under study and will be presented in future work.

We have studied the phenomenologically viable parameter space,
and showed that while our scenario can evade CMB constraints from primordial black hole formation, typically
quite stringent, it is difficult to generate the observed amount of curvature perturbations. An extra ingredient,
like a curvaton field, is likely needed to match the measured power spectrum. 
We find that the promotion of the relaxion to an inflaton can accommodate a cutoff close to the Planck scale, significantly above the one found in the original proposal~\cite{Graham:2015cka}.

\acknowledgments
We would like to thank Tim Cohen, Erik Kuflik, Josh Ruderman, and Yotam Soreq for collaboration at the embryonic stages
of this work. We benefited from a multitude of discussions with P. Agrawal, B. Batell, C. Csaki, P. Draper, S. Enomoto, W. Fischler, R. Flauger, P. Fox,
R. Harnik, A. Hook, K. Howe,  S. Ipek, J. Kearney, H. Kim, G. Marques-Tavares, L. McAllister, M.~McCullough, S. Nussinov, S. Paban, E. Pajer, M.~Peskin, G. Perez, D. Redigolo, A. Romano, R. Sato, L. Sorbo, and M. Takimoto.
This work is supported in part by the I-CORE Program of the Planning Budgeting Committee and the Israel Science Foundation (grant No. 1937/12), by the European Research Council (ERC) under the EU Horizon 2020 Programme (ERC- CoG-2015 - Proposal n. 682676 LDMThExp) and by the German-Israeli Foundation (grant No. I-1283- 303.7/2014).
The work of LU was performed in part at the Aspen Center for Physics, which is supported by National Science Foundation grant PHY-1066293, and was partially supported by a grant from the Simons Foundation.

\appendix 

\section{A clockwork model} \label{app:clock}

A possible UV completion for the model presented in Section~\ref{sec:model} is provided by the clockwork mechanism \cite{Choi:2015fiu, Kaplan:2015fuy}(see also Refs.~\cite{Kim:2004rp, Harigaya:2014eta, Choi:2014rja, Higaki:2014pja, Kappl:2014lra, Ben-Dayan:2014zsa, Bai:2014coa, delaFuente:2014aca,Giudice:2016yja,Craig:2017cda}).
The construction relies on the potential~\cite{Kaplan:2015fuy}
\be \label{clockpot}
V(\Phi) = \sum_{j=1}^{N+1} \left( -\mu^2_\Phi \Phi_j^\dagger \Phi_j + \frac{\lambda_\Phi}{4} | \Phi_j^\dagger \Phi_j |^2 \right)
+ \sum_{j=1}^{N} \left( \epsilon_\Phi \Phi_j^\dagger \Phi^3_{j+1} + {\rm h.c.}  \right) \, ,
\ee
where $\Phi_j$'s are complex scalar fields.
The terms in the first sum respect a global $U(1)^{N+1}$ symmetry, while the second sum explicitly breaks it to a $U(1)$. 
The fields $\Phi_j$ have charges $Q = 1, \frac{1}{3}, \frac{1}{9}, \dots, \frac{1}{3^N}$ under the unbroken $U(1)$.
As $\mu^2_\Phi > 0$, all the $U(1)$'s are spontaneously broken at a scale $f = \sqrt{(2\mu^2_\Phi)/\lambda_\Phi}$. 
The corresponding Nambu-Goldstone bosons (NGB) obtain a mass proportional to $\sqrt{\epsilon_\Phi} f$, 
due to the terms with $\epsilon_\Phi \ll 1$ in Eq.~\eqref{clockpot}, except for 
that associated with the $U(1)$ which is not explicitly broken. 
The latter NGB (massless at this stage) is given by the combination $\phi = {\cal N} \left( 1 \ \frac{1}{3} \ \frac{1}{9} \ \dots \frac{1}{3^N} \right)$, which we identify with the relaxion. Here ${\cal N}$ is a normalization factor. The relaxion has exponentially suppressed
overlap with the operators $\Phi_{N+1}$ couple to.

We couple $\Phi_1$ to fermions charged under a non-Abelian gauge group that confines at the scale $\Lambda_{\rm wig}$.  
Via the one-loop triangle diagram the relaxion obtains the coupling
$\frac{\alpha_1}{8\pi}\frac{\phi}{f} G_1\tilde G_1$,
which gives rise to the periodic wiggle potential. We couple $\Phi_{N+1}$ to fermions charged under another gauge group with confining scale $\Lambda_N \gg \Lambda_{\rm wig}$. Because of the suppressed overlap of the relaxion with the $N+1$ field,
 the operator leads to the coupling $\frac{\alpha_N}{8\pi}\frac{\phi}{F} G_{N+1}\tilde G_{N+1}$, with $F = 3^N f \gg f$.  Below the confining scale, the potential $\Lambda_N^4 \cos \frac{\phi}{F}$, responsible for the rolling, emerges. 
By controlling which of the scalars couple to the dark photon, 
one may control the strength of the photon coupling to the relaxion, 
namely one can set the value of $c_{\gamma_{(D)}}$ over a large range~\cite{Farina:2016tgd}. 
For example, by charging the fermions at the ($j+1$)th site under the Abelian gauge symmetry, the relaxion--photon coupling would be 
$c_{\gamma_{(D)}}\sim \frac{\alpha_{(D)}}{2\pi}3^{-j}$. 

The full clockwork-inspired Lagrangian for the relaxion that we consider is then 
\beqn
-\mathcal{L}(h,\phi) & =  & \frac{1}{2} \Lambda^2 \left( 1  - g_h \,\cos \frac{\phi}{F}\right) h^2 + \frac{1}{4} \lambda h^4 + \frac{m_W^4}{4 \lambda} + V_{\rm roll}(\phi) + V_{\rm wig}(\phi)  \nonumber \\
& {} & + c_{\gamma_{(D)}} \frac{\phi}{4 f} F_{(D)} \tilde F_{(D)} \, ,  \label{fullpot} \\
V_{\rm roll}(\phi) & = & \Lambda_N^4 \left(\alpha_{cc} - \cF \right)  \, , \label{Vrollcl} \\
V_{\rm wig}(\phi) & = & \Lambda_{\rm wig}^4 \cf  \, . \label{Vwigglescl}
\eeqn
Here, $\alpha_{cc}$ is a dimensionless constant that we use to tune the cosmological constant to zero.
To make sense of the notion of pNGB, all the scales corresponding to explicit breaking have to be smaller than the spontaneous breaking scale, so we have the following hierarchy\footnote{In the absence of tuning, the scale $\Lambda_N$ is expected to be of order $\Lambda$ (up to a loop factor), as the $h^2$ term is going to generate $\frac{g_h \Lambda^4}{16 \pi^2} \cF$ anyway.}
\be \label{pNGBcondition}
\Lambda_{\rm wig} \ll \Lambda \sim \Lambda_N \ll f \ll F \, .
\ee
The dimensionless parameter $g_h > 1$ determines the point at which we switch from the unbroken to the broken EW phase:
\be
\cF > \frac{1}{g_h} \qquad {\rm broken \ phase} \, .
\ee
With these conventions, we imagine that the rolling starts from $\phi / F$ between 0 and $\pi$ and rolls down to the left. 
We define $\phi_0$ as the point where $m_h = 0$: 
\be
 \cos \frac{\phi_0}{F}\, =  \,\frac{1}{g_h} \, .
\ee
Expanding around this point, $\phi = \phi_0 + \dphi$, we have
\begin{equation} \label{mu2clock}
\mu^2(\dphi) \,\simeq\,  g_h \Lambda^2  \sFo \,\frac{\dphi}{F} \, .
\end{equation}
Setting $\mu^2(\dphi_{\rm EW}) \simeq - m_W^2$ we find
\be
\frac{\dphi_{\rm EW}}{F}  \simeq  - \frac{m_W^2}{\Lambda^2} \frac{1}{g_h \sFo} \, . \label{varphiEW} 
\ee
We want to tune the cosmological constant at this point:
\be \label{vrollcc}
V_{\rm roll}(\phi) = \Lambda_N^4 \left[ \cos \frac{\phi_0 + \dphi_{\rm EW}}{F} - \cF \right] \, ,
\ee
which after expanding around $\phi_0$ reads
\be
V_{\rm roll}(\dphi) \simeq  \Lambda_N^4 \sFo \left[ \frac{\dphi}{F} - \frac{\dphi_{\rm EW}}{F} \right] \, .
\ee
Putting all the pieces together we have
\beqn 
V(h, \dphi) & = & \frac{1}{2} g_h  \frac{\Lambda^2}{F} \sFo \ \dphi \ h^2 + \frac{1}{4} \lambda h^4 + \frac{m_W^4}{4 \lambda} + V_{\rm roll}(\dphi)+ V_{\rm wig}(\dphi) \, , \label{Vhdphicl} \\
V_{\rm roll}(\dphi) & = &  \frac{\Lambda_N^4}{F} \sFo (\dphi - \dphi_{\rm EW}) \, , \label{Vrdphicl} \\
V_{\rm wig}(\dphi) & = & \Lambda_{\rm wig}^4 \cos \frac{\phi_0 + \dphi}{f} \, . \label{Vrwigcl}
\eeqn
We see that, by identifying
\be
m \equiv  \frac{\Lambda^2}{F} \sFo \sim  \frac{\Lambda_N^2}{F} \sFo \, ,
\ee
we can match this potential to the one given at the end of Section~\ref{sec:model}.


\section{Slow Roll Conditions} \label{sec:slowroll}
\subsection{Regime 1}
In Section~\ref{regime1}, we discussed the slow-roll conditions in regime 1, where the barriers from the wiggles are not yet large,
namely the condition $V'_{\rm roll}+ V'_{\rm wig}>0$ is satisfied.
We saw that the parameter $\epsilon(\dphi) = -\dot H / H^2$ remains smaller than 1 for values of the potential down to $V \sim \Mp V'_{\rm roll}$.
In this appendix, we discuss in detail the other slow-roll parameter:
\be
\eta(\dphi) \equiv \epsilon(\dphi) - \frac{\delta\ddot\phi}{H\delta\dot\phi} \simeq - \frac{\delta\ddot\phi}{H\delta\dot\phi} \, .
\ee
The last equality holds as long as $\epsilon < 1$.

We start from the equation of motion
\begin{align} \label{EOMapp}
\delta\ddot\phi + 3 H \delta\dot\phi + V'_{\rm roll}+ V'_{\rm wig} =0 \ , 
\end{align}
and define the small parameter 
\begin{align}
\vartheta\equiv \left|\frac{V'_{\rm wig} }{V'_{\rm roll}}\right| \, .
\end{align}
In regime 1, $\vartheta$ typically does not grow larger than 0.1.
We expand $\delta\dot\phi$ as
\begin{align} \label{dphiexpansion}
\delta\dot\phi= \delta\dot\phi^{(0)}+\vartheta\delta\dot\phi^{(1)}+{\cal O}(\vartheta^2) \, .
\end{align}
At zeroth order in $\vartheta$, the equation of motion reads
\be\label{phi0eom}
3 H \delta\dot\phi^{(0)} + V'_{\rm roll} =0 \, ,
\ee
where we dropped $\delta\ddot\phi^{(0)}$ because $\eta^{(0)}\simeq \Mp^2 V''_{\rm roll}/V_{\rm roll}=0$. With the
boundary conditions $\delta\phi=0$ and $\delta\dot\phi=\dot\phi_0 \equiv -V'_{\rm roll}/3H$ at $t = 0$, and treating $H$ 
as roughly constant, we have
\begin{align}\label{phi0solution}
\delta\phi^{(0)}(t) =\dot\phi_0 t + {\cal O}(\epsilon) \, .
\end{align}
At first order in $\vartheta$, the equation of motion is
\be
\label{phi1eom}
\vartheta \delta\ddot\phi^{(1)} + 3 H \vartheta\delta\dot\phi^{(1)} + V'_{\rm wig} =0 \ .  
\ee
Substituting the zeroth order solution \eqref{phi0solution} into Eq.~\eqref{phi1eom}, we get

\begin{align}
\vartheta \delta\ddot\phi^{(1)} + 3 H \vartheta\delta\dot\phi^{(1)} 
- \frac{\Lambda_{\rm wig}^4}{f}\sin\frac{\phi_0+\dot\phi_0 t}{f} &=0\ ,
\end{align}
which can be solved analytically: 
\begin{align}
\label{dotphi1sol}
&\vartheta\delta\dot\phi^{(1)}= 
\frac{-\Lambda_{\rm wig}^4 \left\{
\dot\phi_0\left[ \cos\frac{\phi_0+\dot\phi_0 t}{f} -e^{-3Ht}\cos\frac{\phi_0}{f} \right]
-3H f \left[ \sin\frac{\phi_0+\dot\phi_0 t}{f} - e^{-3Ht}\sin\frac{\phi_0}{f} \right] \right\}}{\dot\phi_0^2+(3Hf)^2}  ,
\\
\label{ddotphi1sol}
&\vartheta\delta\ddot\phi^{(1)}=\frac{\Lambda_{\rm wig}^4}{f} 
\frac{\dot\phi_0^2\sin\frac{\phi_0+\dot\phi_0 t}{f}+3H f \dot\phi_0\cos\frac{\phi_0+\dot\phi_0 t}{f} 
-e^{-3Ht}\{3H f \dot\phi_0\cos\frac{\phi_0}{f}-(3Hf)^2\sin\frac{\phi_0}{f}\}
}{\dot\phi_0^2+(3Hf)^2} \ . 
\end{align}

There are two limits to study:
\begin{enumerate}
\item $3H f\gg |\dot\phi_0|$.

This corresponds to the beginning of regime 1, when $V\sim m_W^2 \Lambda^2$.
  We have
\begin{align}
\vartheta\delta\ddot\phi^{(1)}\simeq\frac{\Lambda_{\rm wig}^4}{f} 
\frac{ \dot\phi_0 }{3Hf}\cos\frac{\phi_0+\dot\phi_0 t}{f} \ ,
\end{align}
and 
\begin{align}
\eta \simeq -\frac{\delta\ddot\phi }{ H \delta\dot\phi} 
\simeq  -\frac{\vartheta\delta\ddot\phi^{(1)}}{H\delta\dot\phi^{(0)}} 
 \sim \vartheta \frac{ \dot\phi_0 }{Hf} \cos\frac{\phi_0+\dot\phi_0 t}{f}\ .
\end{align}
Thus $|\eta| \ll 1$, and $\delta\dot\phi$ is constant to a very good approximation. 
\item $3H f\ll |\dot\phi_0|$.

This is the more interesting limit, which corresponds to the end of regime 1. 
We have
\begin{align}
\vartheta\delta\ddot\phi^{(1)} \simeq 
\frac{\Lambda_{\rm wig}^4}{f}\sin\frac{\phi_0+\dot\phi_0 t}{f} \ ,
\end{align}
and
\begin{align}
\eta \simeq -\frac{\delta\ddot\phi }{ H \delta\dot\phi} 
\simeq  -\frac{\vartheta\delta\ddot\phi^{(1)}}{H\delta\dot\phi^{(0)}}
 \simeq  -\frac{V'_{\rm wig} }{V'_{\rm roll}} \sim \vartheta \ .
\end{align}
This proves that $|\eta| < 1$ also in this limit. Note that $\delta\dot\phi$ stays roughly constant because the relaxion does not gain net kinetic energy from the wiggles. Indeed, 
the maximum deviation from $\delta\dot\phi=\dot\phi_0$ can be estimated by taking one period $t_f=2\pi f/\dot\phi_0$, 
\begin{align}
\left| \frac{\vartheta\delta\ddot\phi^{(1)} t_f}{\dot\phi_0} \right|<\vartheta \frac{2\pi Hf}{\dot\phi_0}\ll1 \, .
\end{align}
\end{enumerate}

\subsection{Regime 2}
Let us rewrite the full EOM as
\be \label{EOMexpfull}
\delta \ddot \phi + 3 H \delta \dot\phi + V' = \frac{C_0}{f_\gamma} \frac{H^4}{(\delta\dot\phi/2 f_\gamma H)^4} e^{-\frac{\pi \delta \dot \phi}{H f_\gamma}} \, ,
\ee
where we have defined $f_\gamma \equiv \frac{f}{c_\gamma}$, and $C_0 \simeq 2.4 \times 10^{-4}$. 
Recall that in our conventions $V' > 0$ and $\delta\dot\phi < 0$.
In Section~\ref{regime2} we claimed that in regime 2 the EOM is well approximated by
\be \label{EOMexpnodd}
V'_{\rm roll}(\dphi) \simeq \frac{C_0}{f_\gamma} \frac{H^4}{(\delta\dot\phi/2 f_\gamma H)^4} e^{-\frac{\pi \delta \dot \phi}{H f_\gamma}}
=\frac{C_0}{f_\gamma} \frac{H^4}{\xi^4} e^{-\pi \xi} \, ,
\ee
with the slow-roll conditions satisfied when $f_\gamma < \frac{\Mp}{|\xi|}$. In what follows we justify these statements.

First, note that in Eq.~\eqref{EOMexpnodd} we are keeping only the rolling potential and neglecting the wiggles. 
We check later what happens when we include the wiggles.
The solution to Eq.~\eqref{EOMexpnodd} is obtained by
\be \label{soldphidot}
\xi  \simeq -\frac{1}{\pi} \ln \left[ \frac{\xi^4 f_\gamma V'_{\rm roll}}{C_0 H^4}  \right]\sim \text{const.}\ , \quad
\delta\dot\phi  \simeq -\frac{H f_\gamma}{\pi} \ln \left[ \frac{\xi^4 f_\gamma V'_{\rm roll}}{C_0 H^4}  \right] \, .
\ee
With this we check the following conditions:
\begin{itemize}
\item The kinetic energy is smaller than the potential
\begin{align}
\frac{\frac{1}{2}\delta\dot\phi^2}{V}\simeq\frac{\frac{1}{2}\left(2\xi H f_\gamma \right)^2}{V}\
=\frac{2}{3}\frac{f^2}{c_\gamma^2}\frac{\xi^2}{\Mp^2} <1 \, ,
\end{align}
\item $H$ is slowly varying, that is $\epsilon = -\frac{\dot H}{H^2} < 1$. Using the Friedmann 
equations we can bring $\epsilon$ to the form~\cite{Anber:2009ua}
\be
\epsilon = \frac{1}{2 \Mp^2 H^2} \left[ \delta\dot\phi^2 + \frac{4}{3} \rho_\gamma \right] \simeq 2 \left( \frac{\xi^2 f^2_\gamma}{\Mp^2} + 
\frac{\rho_\gamma}{V} \right) \, .
\ee
The second term in parentheses is smaller than one for $V > 2 \rho_\gamma = 2 V_{\rm RH}$, that is roughly until reheating. 
Then for the first term we have to impose
\be \label{fgammabound}
f_\gamma \ll \frac{\Mp}{|\xi|} \, .
\ee
\item The term $3H \delta\dot\phi$ in Eq.~\eqref{EOMexpfull} is negligible. We have
\be
\frac{|3 H \delta\dot\phi|}{V'_{\rm roll}} \simeq \frac{2 |\xi| V f_\gamma}{\Mp^2 V'_{\rm roll}} < 1 \, ,
\ee
due to $V < V_{\rm switch} = \frac{\Mp^2 V'_{\rm roll}}{2 |\xi| f_\gamma}$, see Eq.~\eqref{Vswitch}.
\item The term $\delta\ddot\phi$ is negligible. Taking the time derivative of Eq.~\eqref{soldphidot} we find
\be \label{phiddot}
\frac{|\delta\ddot\phi|}{V'_{\rm roll}} \simeq \frac{2|\xi|}{3} \left| -\epsilon \frac{f_\gamma V}{\Mp^2 V'_{\rm roll}} - 
\frac{f_\gamma^2}{\pi \Mp^2} \left( \frac{V V''_{\rm roll}}{V'^2_{\rm roll}} -2  \right)  \right| < 1 \, .
\ee
Here, the first term is smaller than one for $V < V_{\rm switch}$, the second vanishes as $V''_{\rm roll} = 0$,
the third is small as long as Eq.~\eqref{fgammabound} is satisfied.
\end{itemize}
We see that the condition of Eq.~\eqref{fgammabound} is enough to guarantee slow-roll in this approximation.
Next we examine what happens when we take into account also the wiggles, the main worry being that
$\delta\ddot\phi$ could grow larger than $V'$ because of the nonzero $V''_{\rm wig}$. 
We proceed as in the previous section, by defining the small parameter 
$\vartheta = \left\vert \frac{V'_{\rm wig}}{V'_{\rm roll}} \right\vert$ and expanding $\delta\dot\phi$ as in Eq.~\eqref{dphiexpansion}.
We have already solved the zeroth order EOM, that is Eq.~\eqref{EOMexpnodd}. At first order in $\vartheta$ the EOM is
\begin{align}
\vartheta \delta\ddot\phi^{(1)} + 3 H \vartheta\delta\dot\phi^{(1)} + V'_{\rm wig} 
&=
2^4 H^8 f_\gamma^3 \left(\frac{e^{-{\pi(\delta\dot\phi^{(0)}+\vartheta\delta\dot\phi^{(1)})}/{H f_\gamma}}}{(\delta\dot\phi^{(0)}+\vartheta\delta\dot\phi^{(1)})^4 }
-\frac{e^{-{\pi \delta\dot\phi^{(0)}}/{H f_\gamma}}}{(\delta\dot\phi^{(0)})^4 } 
\right) \nn
\\
&\simeq V'_{\rm roll}\left(-4-\frac{\pi \delta\dot\phi^{(0)}}{H f_\gamma} \right)\frac{\vartheta\delta\dot\phi^{(1)}}{\delta\dot\phi^{(0)}}
 \ .  \label{phi1eomexp}
\end{align}
We linearized the equation in the second line assuming $\frac{\pi c_\gamma \vartheta\delta\dot\phi^{(1)}}{fH}$ is another small parameter. One can check this assumption is correct after finding the solution. 
Now the photon friction is just a modification of the Hubble friction term, and the EOM reduces to Eq.~\eqref{phi1eom}
with the replacements
\begin{align}
&\dot\phi_0\to \delta\dot\phi^{(0)} = 2\xi H f_\gamma, 
\\
&H\to H_{\rm eff}\equiv H+\frac{V'_{\rm roll}}{3\delta\dot\phi^{(0)}}\left(4+\frac{\pi \delta\dot\phi^{(0)}}{H f_\gamma} \right)
\simeq 2\pi|\xi| H . 
\end{align}
The solution then is that of Eq.~\eqref{dotphi1sol}. 
It is easy to verify that $\frac{\vartheta\delta\dot\phi^{(1)}}{\delta\dot\phi^{(0)}} < 1$ and
$\frac{\pi\vartheta\delta\dot\phi^{(1)}}{H f_\gamma} < 1$, which confirms the consistency of our expansion in Eq.~\eqref{phi1eomexp}.
Again we can check what happens to the acceleration $\vartheta \delta\ddot\phi^{(1)}$ in two limits:
\begin{enumerate}
\item $3H_{\rm eff} f\gg |\delta\dot\phi^{(0)}|$. 

Here we have
\begin{align}
|\vartheta\delta\ddot\phi^{(1)}| \simeq\frac{\Lambda_{\rm wig}^4}{f} 
\frac{ \delta\dot\phi^{(0)} }{3H_{\rm eff} f}\cos\frac{\phi_0+\delta\dot\phi^{(0)} t}{f} \ll V' \ .
\end{align}
\item $3H_{\rm eff} f \ll |\delta\dot\phi^{(0)}|$.

Here we have
\begin{align}
\frac{|\vartheta\delta\ddot\phi^{(1)}|}{V'} \simeq \frac{1}{V'} \frac{\Lambda_{\rm wig}^4}{f} 
\sin\frac{\phi_0+\delta\dot\phi^{(0)} t}{f} \simeq \frac{V'_{\rm wig}}{V'_{\rm roll}} = \vartheta < 1 \ .
\end{align}
\end{enumerate}
We conclude that even when taking into account the wiggle potential, the acceleration $\delta\ddot\phi$ in
Eq.~\eqref{EOMexpfull} remains negligible.

So far we have checked the self-consistency conditions, $\frac{|3 H \delta\dot\phi|}{V'}, \frac{|\delta\ddot\phi|}{V'}<1$, 
based on the solution of Eq.~\eqref{soldphidot}.
At last, we examine a possible $\mathcal{O}(\vartheta^{0})$ correction to  $\frac{|\delta\ddot\phi|}{V'}$. 
This is because one could worry that $\delta\ddot\phi$ is large when we transition
from regime 1 to 2, and perhaps it is not a good approximation to neglect it in the EOM. We show that even if
$\delta\ddot\phi$ is large at the beginning of regime 2, the solution quickly converges to the one of Eq.~\eqref{soldphidot}. 
Keeping the $\delta\ddot\phi$ term in the EOM introduces extra time-dependence of $\delta\dot\phi^{(0)}$ leading to a different contribution to $\frac{|\delta\ddot\phi|}{V'}$. To see this effect, we consider the following differential equation 
\begin{align}
\delta \ddot \phi  + V_{\rm roll}' = \frac{C_0}{f_\gamma} \frac{H^4}{\xi^4} e^{-\frac{\pi \delta \dot \phi}{H f_\gamma}} \, .
\end{align}
For simplicity we drop the $3H\delta\dot\phi$ term, and neglect again the time-dependence of the prefactor on the RHS, keeping only the stronger time-dependence in the exponent. The solution is 
\begin{align}
\delta \dot \phi(t)&=\frac{H f_\gamma}{\pi} \ln \left[ \frac{C_0 H^4}{\xi^4 f_\gamma V'_{\rm roll}} 
+e^{-\frac{\pi V'_{\rm roll}}{f_\gamma H}t} \left(e^{-2\pi |\xi_0|} -\frac{C_0 H^4}{\xi^4 f_\gamma V'_{\rm roll}} \right) \right]
\\
&=\frac{H f_\gamma}{\pi} \ln \left[\frac{C_0 H^4}{\xi^4 f_\gamma V'_{\rm roll}} \left(
1-\frac{\delta\ddot\phi(0)}{\delta\ddot\phi(0)+V'_{\rm roll}} e^{-\frac{\pi V'_{\rm roll}}{f_\gamma H}t} \right)\right] \ ,
\\
\label{phiddot2}
\delta \ddot \phi(t)&=
\frac{\delta\ddot\phi(0)}{\left(\frac{\delta\ddot\phi(0)}{V'_{\rm roll}}+1\right)
e^{\frac{\pi V'_{\rm roll}}{f_\gamma H}t}-\frac{\delta\ddot\phi(0)}{V'_{\rm roll}}}
\end{align}
where $t=0$ corresponds to the time of transition from regime 1 to 2, and $\xi_0\equiv \frac{\delta\dot\phi(0)}{2f_\gamma H}<0$.
At the transition the acceleration could be sizable, $\frac{|\delta\ddot\phi|}{V'} = \mathcal{O}(1)$. The contribution of Eq.~\eqref{phiddot2} is not included in Eq.~\eqref{phiddot}.
However, $\frac{|\delta\ddot\phi|}{V'}$ quickly becomes small  because the exponential damping factor is much larger than $H$, 
\begin{align}
\frac{\pi V'_{\rm roll}}{f_\gamma H}\sim 6\pi|\xi| H \ . \label{largexp}
\end{align}
Here we used Eq.~\eqref{Vswitch}. 
Despite regime 2 only lasts $\cal O$(20) e-foldings, Eq.~\eqref{largexp} implies that $\frac{|\delta\ddot\phi|}{V'}$ becomes negligible
sooner than one e-fold after the transition ($t\gtrsim\frac{1}{6\pi|\xi| H}$).


\section{Thermal effects on gauge-field production} \label{sec:thermal}

At zero temperature, the equation of motion (EOM) for the polarization that gets exponentially enhanced reads
\be \label{EOMAmin}
\frac{\partial^2A_-}{\partial \tau^2} + \left(k^2 + a k c_\gamma \frac{\dot\phi}{f} \right) A_- = 0 \, .
\ee
Here $\tau$ is the conformal time, defined as $a d\tau = dt$, with $t$ the cosmic time. An overdot denotes a derivative with respect to $t$. In our conventions $\dot\phi < 0$.
 Written in terms of $t$, the EOM is
\be \label{EOMAt}
\ddot A_- + H \dot A_- + \left(\frac{k^2}{a^2} +  \frac{k}{a} c_\gamma \frac{\dot\phi}{f} \right) A_- = 0 \, ,
\ee
from which we can read off explicitly the dispersion relation
\be \label{disp0}
\frac{\omega^2}{a^2} = \frac{k^2}{a^2} +  \frac{k}{a} c_\gamma \frac{\dot\phi}{f} \, .
\ee
The mode $A_-$ experiences tachyonic enhancement when $\omega^2 < 0$. 
The easiest way to find $A_-$ that solves Eq.~\eqref{EOMAmin} is via the WKB approximation:
\be \label{WKBsol2}
A_- (k, \tau) \simeq \frac{1}{\sqrt{2 \Omega(k,\tau)}} e^{\int^\tau d\tau' \Omega(k,\tau')} \, ,
\ee
where, $\Omega \equiv i \omega$.
This approximation holds as long as we satisfy the adiabatic condition
\be \label{adiab}
\left\vert \frac{\partial \Omega}{\partial \tau} \frac{1}{\Omega^2} \right\vert \ll 1 \, .
\ee
Then one can compute
\begin{align}
\langle \vec E \cdot \vec B \rangle & = \frac{1}{4\pi^2 a^4} \int dk k^3 \frac{\partial}{\partial \tau} \vert A_-(k,\tau) \vert^2 \, , \label{EdotB0} \\
\frac{1}{2} \langle \vec E^2 \rangle & = \frac{1}{4\pi^2 a^4} \int dk k^2  \left\vert \frac{\partial}{\partial \tau} A_-(k,\tau) \right\vert^2 \, , \label{E20} \\
\frac{1}{2} \langle \vec B^2 \rangle & = \frac{1}{4\pi^2 a^4} \int dk k^2  k^2 \vert  A_-(k,\tau) \vert^2 \,  \label{B20} .
\end{align}

At finite temperature, in the long wavelength limit, the dispersion relation \eqref{disp0} is modified to~\cite{Kapusta:2006pm}
\be \label{dispT}
\frac{\omega^2}{a^2} - \frac{k^2}{a^2} - \frac{k}{a} c_\gamma \frac{\dot \phi}{f} = m^2_D\frac{\omega}{k}
\left[ \frac{\omega}{k} + \frac{1}{2} \left( 1 - \frac{\omega^2}{k^2} \right) \ln \frac{\omega + k}{\omega - k} \right] \, , \quad \frac{k}{a} \ll m_D \, ,
\ee
while in the short wavelength limit it is
\be \label{dispTshort}
\frac{\omega^2}{a^2} - \frac{k^2}{a^2} - \frac{k}{a} c_\gamma \frac{\dot \phi}{f} = m^2_D \, , \quad  \quad \frac{k}{a} \gg m_D \, .
\ee
In these expressions
\be \label{mD}
m_D^2 = \frac{g^2 T^2}{6} \, ,
\ee
with $g$ the $U(1)$ gauge coupling. We want to find tachyonic solutions, $\Omega = i \omega >0$, of the equations above.

\subsection{Long wavelength}
Let's consider Eq.~\eqref{dispT} first.
\begin{itemize}
\item $\Omega \gg k$

The RHS of Eq.~\eqref{dispT} reduces to $\simeq \frac{2}{3} m_D^2$, and the dispersion relation to
\be
\frac{\Omega^2}{a^2} + \frac{k^2}{a^2} + \frac{k}{a} c_\gamma \frac{\dot \phi}{f} + \frac{2}{3}m_D^2 = 0 \, .
\ee
Given $\frac{k}{a} \ll m_D$, we see there is no real and positive solution (no tachyonic modes) for $\Omega$ in this limit. Therefore
we turn to the opposite limit:
\item $\Omega \ll k$

The right hand side term in Eq.~\eqref{dispT} reduces to
\be
m^2_D\frac{\omega}{k} \frac{1}{2} \ln(-1) = \frac{\pi}{2} m^2_D\frac{i \omega}{k} = \frac{\pi}{2} m^2_D\frac{\Omega}{k} \, ,
\ee
so the dispersion relation becomes
\be \label{dispTapp}
\frac{\Omega^2}{a^2} + \frac{k^2}{a^2} + \frac{k}{a} c_\gamma \frac{\dot \phi}{f} + \frac{\pi}{2} m^2_D\frac{\Omega}{k} = 0 \, .
\ee
Using the dimensionless variables $\xi= c_\gamma \frac{\dot\phi}{2Hf} < 0$ and $x = -k\tau > 0$ this equation is
\be
\Omega^2 + k^2 + k^2 \frac{2\xi}{x} + k \frac{\pi}{2} \frac{m_D^2}{H^2} \frac{\Omega}{x^2} = 0 \, ,
\ee
with solution
\be
\Omega = \frac{k}{4x^2} \left( \sqrt{\pi^2 \frac{m_D^4}{H^4} -16 x^4 +32 x^3 |\xi|} -\pi \frac{m_D^2}{H^2}  \right) \, .
\ee
In the cases we are interested in, we have $m_D \gg H$. In this limit the solution simplifies to
\be \label{OmegaTinf}
\Omega = \frac{2}{\pi} \frac{H^2}{m_D^2} k x (2|\xi| - x) \, .
\ee
$\Omega$ is positive (we have tachyonic modes) as long as $x< 2|\xi|$.

The adiabatic condition is
\be 
\left\vert \frac{\partial \Omega}{\partial \tau} \frac{1}{\Omega^2} \right\vert = \left\vert \pi \frac{m_D^2}{H^2} \frac{|\xi|-x}{x^2(2|\xi| - x)^2}  \right\vert< 1 \, .
\ee
This is satisfied only in a very narrow range of $x$ close to $|\xi|$:
\be \label{xint}
|\xi| - \frac{\xi^4}{\pi} \frac{H^2}{m_D^2} < x < |\xi| + \frac{\xi^4}{\pi} \frac{H^2}{m_D^2} \, .
\ee
Now that we have $\Omega$ we can get $A_-$ using again the WKB approximation.
We have
\be
\int d\tau \Omega = \frac{1}{k} \int_{x_{\rm min}}^{x_{\rm max}} dx \Omega \simeq \frac{4}{\pi^2} \frac{H^4}{m_D^4} \xi^6 \, .
\ee
In the denominator of Eq.~\eqref{WKBsol2} we can approximate $\Omega$ with $\Omega(x = |\xi|)$. Then the WKB solution is
\be \label{AminTWKB}
A_-(k,x) \simeq \sqrt{\frac{\pi}{4 k}} \frac{m_D}{H} \frac{1}{|\xi|} e^{\frac{4}{\pi^2} \frac{H^4}{m_D^4} \xi^6} \, .
\ee
With this, we can compute the E and B fields using Eqs.~\eqref{EdotB0}, \eqref{E20}, \eqref{B20}. Since the WKB approximation is valid 
in a very narrow range \eqref{xint}, we estimate the integrals as follows. We first change variable from $k$ to $x$. We estimate 
$dx \simeq 2  \frac{\xi^4}{\pi} \frac{H^2}{m_D^2}$, the width of the interval \eqref{xint}, and we substitute $x = |\xi|$. The results are
\beqn
\langle \vec E \cdot \vec B \rangle & \simeq & \frac{1}{2 \pi^3} \frac{H^2}{m_D^2} H^4 |\xi|^7 e^{\frac{4}{\pi^2} \frac{H^4}{m_D^4} \xi^6} \, , \label{EBT} \\
\frac{1}{2} \langle \vec E^2 \rangle & \simeq & \frac{1}{2 \pi^4} \frac{H^4}{m_D^4} H^4 |\xi|^9 e^{\frac{4}{\pi^2} \frac{H^4}{m_D^4} \xi^6}  \, , \label{ET} \\
\frac{1}{2} \langle \vec B^2 \rangle & \simeq  & \frac{1}{8 \pi^2}  H^4 |\xi|^5 e^{\frac{4}{\pi^2} \frac{H^4}{m_D^4} \xi^6} \label{BT} \, .
\eeqn
Note the E field is suppressed compared to the B field at finite temperature.

\end{itemize}

If $\frac{c_\gamma}{f} \langle \vec E \cdot \vec B \rangle$ grows large enough to become comparable to $V'$, the equation of motion of the inflaton
becomes 
\be
V' \simeq \frac{c_\gamma}{f} \langle \vec E \cdot \vec B \rangle \, .
\ee
In this regime at finite $T$ we have
\be
\frac{4}{\pi^2} \frac{H^4}{m_D^4}|\xi|^6 = \ln \left[ 2\pi^3 \frac{m_D^2}{H^2} \frac{f V'}{c_\gamma H^4} \frac{1}{|\xi|^7} \right] \, ,
\ee
from which we find
\be \label{phidotT}
|\dot\phi| = \frac{2 H f}{c_\gamma} \left( \frac{m_D}{H} \right)^{2/3} \left( \frac{\pi^2}{4} \ln \left[ 2\pi^3 \frac{m_D^2}{H^2} \frac{f V'}{c_\gamma H^4} \frac{1}{|\xi|^7} \right] \right)^{1/6} \, .
\ee
Compared to the zero temperature case, where $|\dot\phi| \propto fH$, we see that at finite temperature the velocity is enhanced by
a factor of $\left( \frac{m_D}{H} \right)^{2/3}$. 
We derived this result assuming inflation, but it holds also in the radiation dominated (R.D.) era. Indeed during R.D. we have
\be
a(t) = \left( \frac{t}{t_0} \right)^{1/2} \, , \quad  H = \frac{1}{2t} \, , \quad  \tau = \int \frac{dt}{a} = \frac{a}{H_0} = \frac{1}{a H} \, ,
\ee
and one can check that we get again Eq.~\eqref{OmegaTinf}, with $x$ now defined without the minus sign, $x \equiv k \tau$, because $\tau$ is positive in R.D. The rest of the derivation then follows. 


\subsection{Short wavelength}
In the short wavelength limit, $m_D \ll \frac{k}{a}$, we can treat $m_D$ as a small perturbation in Eq.~\eqref{dispTshort}. We then find ourselves in a situation similar to the zero temperature case.

\section{Electric field}\label{app:electricfield}
We discuss here some properties of the classical electric field formed by the exponential number of photons. 
 
\subsection{Coherence}
The comoving momentum of photons with the largest tachyonic enhancement is $k_{*}=-|\xi|/\tau$ 
(the physical momentum is $q_{\gamma*} = |\xi| H$). The occupation number is given by the number of photons in  
 the coherent volume (within the de Broglie wavelentgh),  $V_{\rm coh} \sim  (|\xi| aH)^{-3}$,  
\begin{align}
\frac{V_{\rm coh}}{(aH)^{-3}}\omega_{k_*} \left\vert \frac{\partial}{\partial \tau} A_-^{\vec k_*}  \right\vert^2 \sim \frac{e^{2\pi |\xi|}}{|\xi|^3} \gg 1 \, .
\end{align}
This number is significantly larger than 1, implying that the photons are coherent and form a classical field.

\subsection{Size and Direction}

Even if numerous photons are produced, one might wonder if their random directions result in a zero net electric field.
Randomized photons in a microscopic scale must have high momentum, but we have seen that those produced 
exponentially in our model have low momentum, $k_*$, instead. Thus, at a comoving scale larger than $k_*^{-1}$, roughly,
we expect zero electric field, but we will have a non-zero field when we zoom into scales smaller than $k_*^{-1}$.
%

We can make these statements more explicit by using an averaged electric field within a radius $R$, 
\footnote{We thank Masahiro Takimoto for suggesting this quantity.}
 \begin{align}
\vec E_R(t, \vec x) &\equiv  \int_{V_R}\!\!\frac{d^3 x_1}{V_R} \vec{E}(t, \vec x+\vec x_1)
=
\frac{-1}{a^2}\sum_{\lambda} \int_{V_R}\!\!\frac{d^3 x_1}{V_R}\int \frac{d^3k}{\left(2\pi \right)^{3/2}}\left[\vec\epsilon_{\lambda,\vec k}\,  \frac{\partial A_\lambda^{\vec k}}{\partial \tau}\,a_\lambda^{\vec k}\,e^{i\vec k\cdot(\vec x+\vec x_1)}+{\mathrm {h.c.}}\right]
\end{align}
where $V_R$ is volume inside a sphere with the radius $R$ from  $\vec{x}$. We study the dispersion  of the averaged electric field,  
 \begin{align}
\langle {\vec E_R}^2(t, \vec x)\rangle
=&\int \frac{d^3 x}{V} \langle 0| {\vec E}_R^2(x)|0 \rangle
\\
=&
\frac{1}{a^4}\sum_{\lambda} \int_{V_R}\!\! \frac{d^3 x_1}{V_R}\frac{d^3 x_2}{V_R}
\int \frac{d^3k }{\left(2\pi \right)^{3}} 
\left\vert \frac{\partial A_-^{\vec k} }{\partial \tau}  \right\vert^2
 e^{i\vec k\cdot\vec x_1-i\vec k\cdot\vec x_2}
=
\frac{1}{a^4}\int \frac{d^3k }{\left(2\pi \right)^{3}} 
 \left\vert \frac{\partial A_-^{\vec k}}{\partial \tau}   \right\vert^2
f_R(k)
\end{align}
 where  $f_R(k)\equiv \left(\int_{V_R}\frac{d^3 x}{V_R} e^{i\vec k\cdot\vec x} \right)^2=
 {9\left[\sin(kR)-k R\cos(kR)\right]^2}/{ (kR)^6}$. 
Since $f_R(k)$ is a function damping quickly for $k\gg R^{-1}$, for simplicity we treat it as a step function, $f_R(k)\to \Theta(R^{-1} -k)$. Also, we approximate $ \left\vert {\partial A_-^{\vec k}}/{\partial \tau}   \right\vert^2$ as 
\begin{align}
 \left\vert \frac{\partial A_-^{\vec k}}{\partial \tau}   \right\vert^2 \simeq \Theta(|\vec k|-k_{IR})\Theta(2k_*-|\vec k|)\rho( k) \, ,
\end{align}
because the production of non-tachyonic photons, with momentum $k> 2k_*$, is negligible. The IR cutoff is needed because there are no zero momentum photons. We  examine two cases, with microscopic and macroscopic scales $R$,
\begin{align}
\langle \vec E_R^2(t, \vec x)\rangle 
\simeq
\frac{1}{a^4}\int \frac{d^3k }{\left(2\pi \right)^{3}} 
 \Theta(|\vec k|-k_{IR})\Theta(k_{UV}-|\vec k|)\rho( k)
f_R(k)\simeq   
  \begin{cases}
    0 & (R> k_{IR}^{-1}),\\
   \langle E^2\rangle  & (R< (2k_{*})^{-1}).
  \end{cases}
\end{align}
Averaging over large scales $(R> k_{IR}^{-1})$, there is no net electric field, while at small scales $(R< (2k_{*})^{-1})$, there is a strong electric field $\langle \vec E^2\rangle\sim \rho_\gamma$, as given in Eq.~\eqref{rhog}. 
The transition from $\langle \vec E^2\rangle$ to zero, going from small to large scales, is expected to be smooth.

The direction of the electric field can appear as a consequence of quantum fluctuations which grow exponentially. 
Our analytic approach is limited to estimating quadratic quantities, such as $\langle \vec E^2 \rangle$, but cannot probe directions. In order to observe the direction, one needs a lattice simulation, which is beyond the scope of this paper. 
For a similar situation of tachyonic instability,  simulations were performed  in Refs.~\cite{Felder:2000hj, Felder:2001kt}. They studied a potential $V=\frac{\lambda}{4}(\phi^2-v^2)^2$ with a homogenous initial condition in the symmetric phase  ($\phi=0$) and initial quantum fluctuations. Then a tachyonic instability drives the inhomogeneity: some patches have $\phi=v$ and other patches have $\phi=-v$.  The appearance of a direction of the electric field is analogous to this inhomogeneity. 
 

\section{Estimate of curvature perturbations} \label{app:curv}

 The power spectrum from the usual vacuum fluctuations of the inflaton, neglecting the contribution 
from gauge fields, is
\begin{equation} \label{powsp}
{\cal P} = \frac{H^4}{4\pi^2 \dot \phi^2} \, .
\end{equation}
We want to check if at 30 e-folds from the end of inflation ${\cal P}$ can match the observed one, 
${\cal P}_{\rm COBE} = 2.5 \times 10^{-9}$. First ,we need to estimate $H$ and $\dot\phi$ at 
that time. We are in regime 1, with
\begin{equation} \label{phidot}
\dot\phi = - \frac{V'}{3H} \, , \qquad H^2 = \frac{V}{3 M_P^2} \, .
\end{equation}
The number of e-folds $N_1$ in this regime, before we switch to the one dominated by photon backreaction, is
\begin{align}
N_1 = \int_{V_{\rm switch}}^{V_1} dV \frac{H}{V' \dot\phi} = \frac{1}{M_P^2 V'^2} \int_{V_{\rm switch}}^{V_1} dV V = 
 \frac{1}{2 M_P^2 V'^2}(V_1^2 - V_{\rm switch}^2) \, ,
\end{align}
where $V_{\rm switch} = \frac{1}{2|\xi_2|} \frac{c_\gamma}{f} M_P^2 V'$ is the potential when we switch to regime 2. 
Hence we have the potential as a function of $N_1$:
\begin{equation} \label{VC}
V_1^2 = V^2_{\rm switch} + 2 M_P^2 V'^2 N_1  \, .
\end{equation}
With this and eq.~\eqref{phidot} we can estimate
\begin{equation}
{\cal P} \simeq 10^{-2} \frac{V_C^3}{M_P^6 V'^2} \simeq 10^{-3}  \frac{V'}{M_P^3} \left[ \frac{1}{\xi_2^2} \left(\frac{c_\gamma \Mp}{f}  \right)^2 + 8 N_1 \right]^{3/2} \, .
\end{equation}
From the bound of eq.~\eqref{fovawindow} we have $\frac{1}{|\xi_2|} \frac{c_\gamma \Mp}{f} < 10$. Also, we consider 
$N_1 \sim {\cal O}(10)$, so the number inside the squared parentheses is of order 100.
Then, considering only the linear slope, we have $V' \simeq m\Lambda^2 \simeq \frac{\Lambda_{\rm wig}^4}{f}$, with the conditions 
$\Lambda_{\rm wig} < m_W < f < M_P$. Thus
\begin{equation}
{\cal P} \simeq 10^{-3} [10^2]^{3/2} \frac{\Lambda_{\rm wig}^3}{M_P^3} \frac{\Lambda_{\rm wig}}{f} <   \frac{m_W^3}{M_P^3} \frac{m_W}{f}  \simeq 10^{-48} \frac{m_W}{f} \ll {\cal P}_{\rm COBE}  \, .
\end{equation}
The curvature perturbations generated only by the linear slope in our model are many orders of magnitude below
what is measured.

\newpage

\bibliographystyle{JHEP}
\bibliography{Relax}

\providecommand{\href}[2]{#2}\begingroup\raggedright\begin{thebibliography}{10}

\bibitem{Graham:2015cka}
P.~W. Graham, D.~E. Kaplan and S.~Rajendran, \emph{{Cosmological Relaxation of
  the Electroweak Scale}},
  \href{http://dx.doi.org/10.1103/PhysRevLett.115.221801}{\emph{Phys. Rev.
  Lett.} {\bf 115} (2015) 221801}, [\href{http://arxiv.org/abs/1504.07551}{{\tt
  1504.07551}}].

\bibitem{Espinosa:2015eda}
J.~R. Espinosa, C.~Grojean, G.~Panico, A.~Pomarol, O.~Pujolàs and G.~Servant,
  \emph{{Cosmological Higgs-Axion Interplay for a Naturally Small Electroweak
  Scale}}, \href{http://dx.doi.org/10.1103/PhysRevLett.115.251803}{\emph{Phys.
  Rev. Lett.} {\bf 115} (2015) 251803},
  [\href{http://arxiv.org/abs/1506.09217}{{\tt 1506.09217}}].

\bibitem{Hardy:2015laa}
E.~Hardy, \emph{{Electroweak relaxation from finite temperature}},
  \href{http://arxiv.org/abs/1507.07525}{{\tt 1507.07525}}.

\bibitem{Jaeckel:2015txa}
J.~Jaeckel, V.~M. Mehta and L.~T. Witkowski, \emph{{Musings on cosmological
  relaxation and the hierarchy problem}},
  \href{http://dx.doi.org/10.1103/PhysRevD.93.063522}{\emph{Phys. Rev.} {\bf
  D93} (2016) 063522}, [\href{http://arxiv.org/abs/1508.03321}{{\tt
  1508.03321}}].

\bibitem{Gupta:2015uea}
R.~S. Gupta, Z.~Komargodski, G.~Perez and L.~Ubaldi, \emph{{Is the Relaxion an
  Axion?}}, \href{http://dx.doi.org/10.1007/JHEP02(2016)166}{\emph{JHEP} {\bf
  02} (2016) 166}, [\href{http://arxiv.org/abs/1509.00047}{{\tt 1509.00047}}].

\bibitem{Batell:2015fma}
B.~Batell, G.~F. Giudice and M.~McCullough, \emph{{Natural Heavy
  Supersymmetry}}, \href{http://dx.doi.org/10.1007/JHEP12(2015)162}{\emph{JHEP}
  {\bf 12} (2015) 162}, [\href{http://arxiv.org/abs/1509.00834}{{\tt
  1509.00834}}].

\bibitem{Matsedonskyi:2015xta}
O.~Matsedonskyi, \emph{{Mirror Cosmological Relaxation of the Electroweak
  Scale}}, \href{http://dx.doi.org/10.1007/JHEP01(2016)063}{\emph{JHEP} {\bf
  01} (2016) 063}, [\href{http://arxiv.org/abs/1509.03583}{{\tt 1509.03583}}].

\bibitem{Marzola:2015dia}
L.~Marzola and M.~Raidal, \emph{{Natural relaxation}},
  \href{http://dx.doi.org/10.1142/S0217732316502151}{\emph{Mod. Phys. Lett.}
  {\bf A31} (2016) 1650215}, [\href{http://arxiv.org/abs/1510.00710}{{\tt
  1510.00710}}].

\bibitem{Choi:2015fiu}
K.~Choi and S.~H. Im, \emph{{Realizing the relaxion from multiple axions and
  its UV completion with high scale supersymmetry}},
  \href{http://dx.doi.org/10.1007/JHEP01(2016)149}{\emph{JHEP} {\bf 01} (2016)
  149}, [\href{http://arxiv.org/abs/1511.00132}{{\tt 1511.00132}}].

\bibitem{Kaplan:2015fuy}
D.~E. Kaplan and R.~Rattazzi, \emph{{A Clockwork Axion}},
  \href{http://arxiv.org/abs/1511.01827}{{\tt 1511.01827}}.

\bibitem{DiChiara:2015euo}
S.~Di~Chiara, K.~Kannike, L.~Marzola, A.~Racioppi, M.~Raidal and C.~Spethmann,
  \emph{{Relaxion Cosmology and the Price of Fine-Tuning}},
  \href{http://dx.doi.org/10.1103/PhysRevD.93.103527}{\emph{Phys. Rev.} {\bf
  D93} (2016) 103527}, [\href{http://arxiv.org/abs/1511.02858}{{\tt
  1511.02858}}].

\bibitem{Ibanez:2015fcv}
L.~E. Ibanez, M.~Montero, A.~Uranga and I.~Valenzuela, \emph{{Relaxion
  Monodromy and the Weak Gravity Conjecture}},
  \href{http://dx.doi.org/10.1007/JHEP04(2016)020}{\emph{JHEP} {\bf 04} (2016)
  020}, [\href{http://arxiv.org/abs/1512.00025}{{\tt 1512.00025}}].

\bibitem{Hebecker:2015zss}
A.~Hebecker, F.~Rompineve and A.~Westphal, \emph{{Axion Monodromy and the Weak
  Gravity Conjecture}},
  \href{http://dx.doi.org/10.1007/JHEP04(2016)157}{\emph{JHEP} {\bf 04} (2016)
  157}, [\href{http://arxiv.org/abs/1512.03768}{{\tt 1512.03768}}].

\bibitem{Fonseca:2016eoo}
N.~Fonseca, L.~de~Lima, C.~S. Machado and R.~D. Matheus, \emph{{Large field
  excursions from a few site relaxion model}},
  \href{http://dx.doi.org/10.1103/PhysRevD.94.015010}{\emph{Phys. Rev.} {\bf
  D94} (2016) 015010}, [\href{http://arxiv.org/abs/1601.07183}{{\tt
  1601.07183}}].

\bibitem{Fowlie:2016jlx}
A.~Fowlie, C.~Balazs, G.~White, L.~Marzola and M.~Raidal, \emph{{Naturalness of
  the relaxion mechanism}},
  \href{http://dx.doi.org/10.1007/JHEP08(2016)100}{\emph{JHEP} {\bf 08} (2016)
  100}, [\href{http://arxiv.org/abs/1602.03889}{{\tt 1602.03889}}].

\bibitem{Evans:2016htp}
J.~L. Evans, T.~Gherghetta, N.~Nagata and Z.~Thomas, \emph{{Naturalizing
  Supersymmetry with a Two-Field Relaxion Mechanism}},
  \href{http://dx.doi.org/10.1007/JHEP09(2016)150}{\emph{JHEP} {\bf 09} (2016)
  150}, [\href{http://arxiv.org/abs/1602.04812}{{\tt 1602.04812}}].

\bibitem{Huang:2016dhp}
F.~P. Huang, Y.~Cai, H.~Li and X.~Zhang, \emph{{A possible interpretation of
  the Higgs mass by the cosmological attractive relaxion}},
  \href{http://dx.doi.org/10.1088/1674-1137/40/11/113103}{\emph{Chin. Phys.}
  {\bf C40} (2016) 113103}, [\href{http://arxiv.org/abs/1605.03120}{{\tt
  1605.03120}}].

\bibitem{Kobayashi:2016bue}
T.~Kobayashi, O.~Seto, T.~Shimomura and Y.~Urakawa, \emph{{Relaxion window}},
  \href{http://arxiv.org/abs/1605.06908}{{\tt 1605.06908}}.

\bibitem{Hook:2016mqo}
A.~Hook and G.~Marques-Tavares, \emph{{Relaxation from particle production}},
  \href{http://dx.doi.org/10.1007/JHEP12(2016)101}{\emph{JHEP} {\bf 12} (2016)
  101}, [\href{http://arxiv.org/abs/1607.01786}{{\tt 1607.01786}}].

\bibitem{Higaki:2016cqb}
T.~Higaki, N.~Takeda and Y.~Yamada, \emph{{Cosmological relaxation and high
  scale inflation}},
  \href{http://dx.doi.org/10.1103/PhysRevD.95.015009}{\emph{Phys. Rev.} {\bf
  D95} (2017) 015009}, [\href{http://arxiv.org/abs/1607.06551}{{\tt
  1607.06551}}].

\bibitem{Choi:2016luu}
K.~Choi and S.~H. Im, \emph{{Constraints on Relaxion Windows}},
  \href{http://arxiv.org/abs/1610.00680}{{\tt 1610.00680}}.

\bibitem{Flacke:2016szy}
T.~Flacke, C.~Frugiuele, E.~Fuchs, R.~S. Gupta and G.~Perez,
  \emph{{Phenomenology of Relaxion-Higgs Mixing}},
  \href{http://arxiv.org/abs/1610.02025}{{\tt 1610.02025}}.

\bibitem{McAllister:2016vzi}
L.~McAllister, P.~Schwaller, G.~Servant, J.~Stout and A.~Westphal,
  \emph{{Runaway Relaxion Monodromy}},
  \href{http://arxiv.org/abs/1610.05320}{{\tt 1610.05320}}.

\bibitem{Choi:2016kke}
K.~Choi, H.~Kim and T.~Sekiguchi, \emph{{Dynamics of cosmological relaxation
  after reheating}},  \href{http://arxiv.org/abs/1611.08569}{{\tt 1611.08569}}.

\bibitem{Lalak:2016mbv}
Z.~Lalak and A.~Markiewicz, \emph{{Dynamical relaxation in 2HDM models}},
  \href{http://arxiv.org/abs/1612.09128}{{\tt 1612.09128}}.

\bibitem{You:2017kah}
T.~You, \emph{{A Dynamical Weak Scale from Inflation}},
  \href{http://arxiv.org/abs/1701.09167}{{\tt 1701.09167}}.

\bibitem{Evans:2017bjs}
J.~L. Evans, T.~Gherghetta, N.~Nagata and M.~Peloso, \emph{{Low-Scale D-term
  Inflation and the Relaxion}},  \href{http://arxiv.org/abs/1704.03695}{{\tt
  1704.03695}}.

\bibitem{Batell:2017kho}
B.~Batell, M.~A. Fedderke and L.-T. Wang, \emph{{Relaxation of the Composite
  Higgs Little Hierarchy}},  \href{http://arxiv.org/abs/1705.09666}{{\tt
  1705.09666}}.

\bibitem{Beauchesne:2017ukw}
H.~Beauchesne, E.~Bertuzzo and G.~Grilli~di Cortona, \emph{{Constraints on the
  relaxion mechanism with strongly interacting vector-fermions}},
  \href{http://arxiv.org/abs/1705.06325}{{\tt 1705.06325}}.

\bibitem{Tangarife:2017vnd}
W.~Tangarife, K.~Tobioka, L.~Ubaldi and T.~Volansky, \emph{{Relaxed
  Inflation}},  \href{http://arxiv.org/abs/1706.00438}{{\tt 1706.00438}}.

\bibitem{Anber:2009ua}
M.~M. Anber and L.~Sorbo, \emph{{Naturally inflating on steep potentials
  through electromagnetic dissipation}},
  \href{http://dx.doi.org/10.1103/PhysRevD.81.043534}{\emph{Phys. Rev.} {\bf
  D81} (2010) 043534}, [\href{http://arxiv.org/abs/0908.4089}{{\tt
  0908.4089}}].

\bibitem{Barnaby:2011vw}
N.~Barnaby, R.~Namba and M.~Peloso, \emph{{Phenomenology of a Pseudo-Scalar
  Inflaton: Naturally Large Nongaussianity}},
  \href{http://dx.doi.org/10.1088/1475-7516/2011/04/009}{\emph{JCAP} {\bf 1104}
  (2011) 009}, [\href{http://arxiv.org/abs/1102.4333}{{\tt 1102.4333}}].

\bibitem{Linde:2012bt}
A.~Linde, S.~Mooij and E.~Pajer, \emph{{Gauge Field Production in Supergravity
  Inflation: Local Non-Gaussianity and Primordial Black Holes}},
  \href{http://dx.doi.org/10.1103/PhysRevD.87.103506}{\emph{Phys. Rev.} {\bf
  D87} (2013) 103506}, [\href{http://arxiv.org/abs/1212.1693}{{\tt
  1212.1693}}].

\bibitem{Pajer:2013fsa}
E.~Pajer and M.~Peloso, \emph{{A review of Axion Inflation in the era of
  Planck}},
  \href{http://dx.doi.org/10.1088/0264-9381/30/21/214002}{\emph{Class. Quant.
  Grav.} {\bf 30} (2013) 214002}, [\href{http://arxiv.org/abs/1305.3557}{{\tt
  1305.3557}}].

\bibitem{Ferreira:2017lnd}
R.~Z. Ferreira and A.~Notari, \emph{{Thermalized Axion Inflation}},
  \href{http://arxiv.org/abs/1706.00373}{{\tt 1706.00373}}.

\bibitem{Heisenberg:1935qt}
W.~Heisenberg and H.~Euler, \emph{{Consequences of Dirac's Theory of
  Positrons}}, \href{http://dx.doi.org/10.1007/BF01343663}{\emph{Z. Phys.} {\bf
  98} (1936) 714--732}, [\href{http://arxiv.org/abs/physics/0605038}{{\tt
  physics/0605038}}].

\bibitem{Schwinger:1951nm}
J.~S. Schwinger, \emph{{On Gauge Invariance and Vacuum Polarization}},
  \href{http://dx.doi.org/10.1103/PhysRev.82.664}{\emph{Phys. Rev.} {\bf 82}
  (1951) 664--679}.

\bibitem{Adshead:2015pva}
P.~Adshead, J.~T. Giblin, T.~R. Scully and E.~I. Sfakianakis,
  \emph{{Gauge-Preheating and the End of Axion Inflation}},
  \href{http://dx.doi.org/10.1088/1475-7516/2015/12/034}{\emph{JCAP} {\bf 1512}
  (2015) 034}, [\href{http://arxiv.org/abs/1502.06506}{{\tt 1502.06506}}].

\bibitem{Sorbo:2011rz}
L.~Sorbo, \emph{{Parity violation in the Cosmic Microwave Background from a
  pseudoscalar inflaton}},
  \href{http://dx.doi.org/10.1088/1475-7516/2011/06/003}{\emph{JCAP} {\bf 1106}
  (2011) 003}, [\href{http://arxiv.org/abs/1101.1525}{{\tt 1101.1525}}].

\bibitem{Abbott:1984qf}
L.~F. Abbott, \emph{{A Mechanism for Reducing the Value of the Cosmological
  Constant}}, \href{http://dx.doi.org/10.1016/0370-2693(85)90459-9}{\emph{Phys.
  Lett.} {\bf B150} (1985) 427--430}.

\bibitem{Giudice:2016yja}
G.~F. Giudice and M.~McCullough, \emph{{A Clockwork Theory}},
  \href{http://arxiv.org/abs/1610.07962}{{\tt 1610.07962}}.

\bibitem{Felder:2000hj}
G.~N. Felder, J.~Garcia-Bellido, P.~B. Greene, L.~Kofman, A.~D. Linde and
  I.~Tkachev, \emph{{Dynamics of symmetry breaking and tachyonic preheating}},
  \href{http://dx.doi.org/10.1103/PhysRevLett.87.011601}{\emph{Phys. Rev.
  Lett.} {\bf 87} (2001) 011601},
  [\href{http://arxiv.org/abs/hep-ph/0012142}{{\tt hep-ph/0012142}}].

\bibitem{Felder:2001kt}
G.~N. Felder, L.~Kofman and A.~D. Linde, \emph{{Tachyonic instability and
  dynamics of spontaneous symmetry breaking}},
  \href{http://dx.doi.org/10.1103/PhysRevD.64.123517}{\emph{Phys. Rev.} {\bf
  D64} (2001) 123517}, [\href{http://arxiv.org/abs/hep-th/0106179}{{\tt
  hep-th/0106179}}].

\bibitem{Copeland:2002ku}
E.~J. Copeland, S.~Pascoli and A.~Rajantie, \emph{{Dynamics of tachyonic
  preheating after hybrid inflation}},
  \href{http://dx.doi.org/10.1103/PhysRevD.65.103517}{\emph{Phys. Rev.} {\bf
  D65} (2002) 103517}, [\href{http://arxiv.org/abs/hep-ph/0202031}{{\tt
  hep-ph/0202031}}].

\bibitem{GarciaBellido:2002aj}
J.~Garcia-Bellido, M.~Garcia~Perez and A.~Gonzalez-Arroyo, \emph{{Symmetry
  breaking and false vacuum decay after hybrid inflation}},
  \href{http://dx.doi.org/10.1103/PhysRevD.67.103501}{\emph{Phys. Rev.} {\bf
  D67} (2003) 103501}, [\href{http://arxiv.org/abs/hep-ph/0208228}{{\tt
  hep-ph/0208228}}].

\bibitem{Cohen:2008wz}
T.~D. Cohen and D.~A. McGady, \emph{{The Schwinger mechanism revisited}},
  \href{http://dx.doi.org/10.1103/PhysRevD.78.036008}{\emph{Phys. Rev.} {\bf
  D78} (2008) 036008}, [\href{http://arxiv.org/abs/0807.1117}{{\tt
  0807.1117}}].

\bibitem{Meerburg:2012id}
P.~D. Meerburg and E.~Pajer, \emph{{Observational Constraints on Gauge Field
  Production in Axion Inflation}},
  \href{http://dx.doi.org/10.1088/1475-7516/2013/02/017}{\emph{JCAP} {\bf 1302}
  (2013) 017}, [\href{http://arxiv.org/abs/1203.6076}{{\tt 1203.6076}}].

\bibitem{Notari:2016npn}
A.~Notari and K.~Tywoniuk, \emph{{Dissipative Axial Inflation}},
  \href{http://arxiv.org/abs/1608.06223}{{\tt 1608.06223}}.

\bibitem{Garcia-Bellido:2016dkw}
J.~Garc{\'\i a-}Bellido, M.~Peloso and C.~Unal, \emph{{Gravitational Waves at
  Interferometer Scales and Primordial Black Holes in Axion Inflation}},
  \href{http://arxiv.org/abs/1610.03763}{{\tt 1610.03763}}.

\bibitem{Liddle:1993fq}
A.~R. Liddle and D.~H. Lyth, \emph{{The Cold Dark Matter Density
  Perturbation}},
  \href{http://dx.doi.org/10.1016/0370-1573(93)90114-S}{\emph{Phys. Rept.} {\bf
  231} (1993) 1--105}, [\href{http://arxiv.org/abs/astro-ph/9303019}{{\tt
  astro-ph/9303019}}].

\bibitem{Flauger:2009ab}
R.~Flauger, L.~McAllister, E.~Pajer, A.~Westphal and G.~Xu, \emph{{Oscillations
  in the CMB from Axion Monodromy Inflation}},
  \href{http://dx.doi.org/10.1088/1475-7516/2010/06/009}{\emph{JCAP} {\bf 1006}
  (2010) 009}, [\href{http://arxiv.org/abs/0907.2916}{{\tt 0907.2916}}].

\bibitem{Piazza:2010ye}
F.~Piazza and M.~Pospelov, \emph{{Sub-eV scalar dark matter through the
  super-renormalizable Higgs portal}},
  \href{http://dx.doi.org/10.1103/PhysRevD.82.043533}{\emph{Phys. Rev.} {\bf
  D82} (2010) 043533}, [\href{http://arxiv.org/abs/1003.2313}{{\tt
  1003.2313}}].

\bibitem{Harnik:2012ni}
R.~Harnik, J.~Kopp and P.~A.~N. Machado, \emph{{Exploring nu Signals in Dark
  Matter Detectors}},
  \href{http://dx.doi.org/10.1088/1475-7516/2012/07/026}{\emph{JCAP} {\bf 1207}
  (2012) 026}, [\href{http://arxiv.org/abs/1202.6073}{{\tt 1202.6073}}].

\bibitem{Kim:2004rp}
J.~E. Kim, H.~P. Nilles and M.~Peloso, \emph{{Completing natural inflation}},
  \href{http://dx.doi.org/10.1088/1475-7516/2005/01/005}{\emph{JCAP} {\bf 0501}
  (2005) 005}, [\href{http://arxiv.org/abs/hep-ph/0409138}{{\tt
  hep-ph/0409138}}].

\bibitem{Harigaya:2014eta}
K.~Harigaya and M.~Ibe, \emph{{Simple realization of inflaton potential on a
  Riemann surface}},
  \href{http://dx.doi.org/10.1016/j.physletb.2014.09.061}{\emph{Phys. Lett.}
  {\bf B738} (2014) 301--304}, [\href{http://arxiv.org/abs/1404.3511}{{\tt
  1404.3511}}].

\bibitem{Choi:2014rja}
K.~Choi, H.~Kim and S.~Yun, \emph{{Natural inflation with multiple
  sub-Planckian axions}},
  \href{http://dx.doi.org/10.1103/PhysRevD.90.023545}{\emph{Phys. Rev.} {\bf
  D90} (2014) 023545}, [\href{http://arxiv.org/abs/1404.6209}{{\tt
  1404.6209}}].

\bibitem{Higaki:2014pja}
T.~Higaki and F.~Takahashi, \emph{{Natural and Multi-Natural Inflation in Axion
  Landscape}}, \href{http://dx.doi.org/10.1007/JHEP07(2014)074}{\emph{JHEP}
  {\bf 07} (2014) 074}, [\href{http://arxiv.org/abs/1404.6923}{{\tt
  1404.6923}}].

\bibitem{Kappl:2014lra}
R.~Kappl, S.~Krippendorf and H.~P. Nilles, \emph{{Aligned Natural Inflation:
  Monodromies of two Axions}},
  \href{http://dx.doi.org/10.1016/j.physletb.2014.08.045}{\emph{Phys. Lett.}
  {\bf B737} (2014) 124--128}, [\href{http://arxiv.org/abs/1404.7127}{{\tt
  1404.7127}}].

\bibitem{Ben-Dayan:2014zsa}
I.~Ben-Dayan, F.~G. Pedro and A.~Westphal, \emph{{Hierarchical Axion
  Inflation}},
  \href{http://dx.doi.org/10.1103/PhysRevLett.113.261301}{\emph{Phys. Rev.
  Lett.} {\bf 113} (2014) 261301}, [\href{http://arxiv.org/abs/1404.7773}{{\tt
  1404.7773}}].

\bibitem{Bai:2014coa}
Y.~Bai and B.~A. Stefanek, \emph{{Natural millicharged inflation}},
  \href{http://dx.doi.org/10.1103/PhysRevD.91.096012}{\emph{Phys. Rev.} {\bf
  D91} (2015) 096012}, [\href{http://arxiv.org/abs/1405.6720}{{\tt
  1405.6720}}].

\bibitem{delaFuente:2014aca}
A.~de~la Fuente, P.~Saraswat and R.~Sundrum, \emph{{Natural Inflation and
  Quantum Gravity}},
  \href{http://dx.doi.org/10.1103/PhysRevLett.114.151303}{\emph{Phys. Rev.
  Lett.} {\bf 114} (2015) 151303}, [\href{http://arxiv.org/abs/1412.3457}{{\tt
  1412.3457}}].

\bibitem{Craig:2017cda}
N.~Craig, I.~Garcia~Garcia and D.~Sutherland, \emph{{Disassembling the
  Clockwork Mechanism}},  \href{http://arxiv.org/abs/1704.07831}{{\tt
  1704.07831}}.

\bibitem{Farina:2016tgd}
M.~Farina, D.~Pappadopulo, F.~Rompineve and A.~Tesi, \emph{{The photo-philic
  QCD axion}}, \href{http://dx.doi.org/10.1007/JHEP01(2017)095}{\emph{JHEP}
  {\bf 01} (2017) 095}, [\href{http://arxiv.org/abs/1611.09855}{{\tt
  1611.09855}}].

\bibitem{Kapusta:2006pm}
J.~I. Kapusta and C.~Gale, \emph{{Finite-temperature field theory: Principles
  and applications}}.
\newblock Cambridge University Press, 2011.

\end{thebibliography}\endgroup

\end{document}